\documentclass[11pt,onecolumn]{IEEEtran}

\ifCLASSINFOpdf
  \usepackage[pdftex]{graphicx}
 \else
  \usepackage[dvips]{graphicx}
\fi

\usepackage[cmex10]{amsmath}
\usepackage{amssymb}

\usepackage{array}

\usepackage{amssymb}
\usepackage{enumitem}
\usepackage{verbatim}
\usepackage{bbold}

\newtheorem{theorem}{Theorem}
\newtheorem{remark}{Remark}

\newtheorem{lemma}{Lemma}
\newtheorem{proposition}{Proposition}

\newcommand{\E}{\mathbb{E}}
\renewcommand{\mod}{\mathrm{mod}}

\newcommand{\beq}{\begin{equation}}
\newcommand{\eeq}{\end{equation}}
\newcommand{\var}{\text{Var}}

\newcommand{\bv}{\mathbf{v}}
\newcommand{\bx}{\mathbf{x}}
\newcommand{\by}{\mathbf{y}}

\newcommand{\bard}{\bar{d}}
\newcommand{\bzero}{\mathbf{0}}

\makeatletter

\newcommand{\Rmnum}[1]{\expandafter\@slowromancap\romannumeral #1@}
\makeatother
\usepackage{mathtools}
\DeclarePairedDelimiter\ceil{\lceil}{\rceil}

\begin{document}
%
\title{Parity Queries for Binary Classification}

\author{Hye Won Chung$^*$,~Ji Oon Lee,~Doyeon Kim~and Alfred O. Hero
\thanks{Hye Won Chung$^*$ (hwchung@kaist.ac.kr) and Doyeon Kim(highlowzz@kaist.ac.kr) are with the School of Electrical Engineering at KAIST in South Korea. Ji Oon Lee (jioon.lee@kaist.edu) is with the Department of Mathematical Sciences at KAIST in South Korea. Alfred O. Hero (hero@eecs.umich.edu) is with the Department of EECS at the University of Michigan. Hye Won Chung was partially supported by National Research Foundation of Korea under grant number 2017R1E1A1A01076340 and by the Ministry of Science and ICT, Korea, under an ITRC Program, IITP-2019-2018-0-01402. Alfred Hero was partially supported by United States Army Research Office under grant W911NF-15-1-0479. This research was presented in part at 2018 IEEE International Symposium on Information Theory in Vail, USA~\cite{chung2018data} and at 2018 56th Annual Allerton Conference on Communication, Control, and Computing (Allerton) \cite{chung2018trade}.  
}}

\maketitle


\begin{abstract}

Consider a query-based data acquisition problem that aims to recover the values of $k$ binary variables from parity (XOR) measurements of chosen subsets of the variables.
Assume the response model where only a randomly selected subset of the measurements is received. 
We propose a method for designing a sequence of queries so that the variables can be identified with high probability using as few ($n$) measurements as possible. We define the query difficulty $\bar{d}$ as the average size of the query subsets and the sample complexity $n$ as the minimum number of measurements required to attain a given recovery accuracy. We obtain fundamental trade-offs between recovery accuracy, query difficulty, and sample complexity. In particular, the necessary and sufficient sample complexity required for recovering all $k$ variables with high probability is $n = c_0 \max\{k, (k \log k)/\bar{d}\}$ and the sample complexity for recovering a fixed proportion $(1-\delta)k$ of the variables for $\delta=o(1)$ is $n = c_1\max\{k, (k \log(1/\delta))/\bar{d}\}$, where $c_0, c_1>0$. 


\end{abstract}


%




\IEEEpeerreviewmaketitle

\section{Introduction}\label{sec:intro}

We consider a task of learning the values of $k$ variables $x_i\in\{0,1\}$, $i=1,\dots, k$, by designing a sequence of queries and receiving responses to a randomly selected subset of the designed queries. 
Such a query-based data acquisition and information recovery problem is one of the fundamental topics in data science and machine learning with diverse applications in crowdsourcing~\cite{karger2014budget, bernstein2011crowds}, active learning~\cite{mackay1992information,settles2010active}, experimental design~\cite{lindley1956measure,fedorov1972theory}, and community recovery or clustering in graphs~\cite{abbe2015community,hajek2017information}. 
For example, in a crowdsourcing system, we can consider the task of classifying $k$ objects in a database, e.g., a set of images, into two groups depending on the binary attributes of the objects, e.g., object's suitability for children. 
Workers in the system are given simple queries, e.g., binary queries, about a chosen subset of objects. Workers who are unsure about their answers may decline to respond by skipping the query. We can model the worker's decision either to answer or not to answer a query as random.
As another example, in clustering problems, the goal is to classify $k$ nodes having binary attributes by observing some similarity measurements for randomly chosen (queried) subsets of the nodes. In these problems, the main goal is to design a sequence of queries to recover the $k$ variables with high probability, at the minimum number $n$ of measurements.
Moreover, we aim to guarantee the recovery with high probability, regardless of which subset of the queries was answered, as long as the received number $n$ of measurements exceeds the minimum threshold.

To model practical data acquisition setups in digital systems we assume that we are allowed to ask a simple parity query, which asks modulo-2 sum of a chosen small subset of the variables, at each querying. The subset size $d$, called query degree, can be varying over queries, but the average size of the subsets is set to be $\bard\in[1,k]$, which we call query difficulty.
The proper query difficulty the query designer can choose might be determined by the applications, depending on the average number of variables answerers can access at each querying or on the average computational ability.
As examples of the applications, we can consider supervised clustering in graphs~\cite{tsourakakis2017predicting} or locally-encodable coding~\cite{mazumdar2017semisupervised}. 
In classical unsupervised community detection problem for the celebrated stochastic block model (SBM)~\cite{holland1983stochastic} or for the censored block model (CBM)~\cite{abbe2013conditional}, the graph data provides a collection of pairwise similarity measurements ($d=2$), which are represented by the presence or absence of an edge between two nodes in the graph. 
Observing whether or not an edge exists between two nodes is equivalent to measuring the parity between the labels of the node pair, possibly corrupted by noise.
To model multi-way interactions among nodes in many real-world problems, hypergraphs, where an edge can connect more than two nodes ($d>2$), have been considered and the clustering in hypergraphs has been studied with applications of social networks, VLSI CAD, molecular biology, etc.~\cite{zhou2007learning, agarwal2006higher, karypis2000multilevel}. 
In the supervised version of the community detection problem, it is assumed that we are allowed to query a simple function, e.g., binary (AND, OR, or XOR) queries, for any small subset of the nodes and the goal is to predict the measurement for unobserved edges or to recover the labels of the nodes~\cite{leskovec2010predicting,tsourakakis2017predicting}. In these problems, the query difficulty might be determined by how many nodes each answerer can access at each querying.
The problem we consider also has relations to the locally encodable coding~\cite{mazumdar2017semisupervised}: a data compression/transmission problem where each coded bit depends only on a small number of input bits.

Two main questions addressed in this paper are: (1) what is the fundamental (information-theoretic) limit on the required sample complexity $n$ to recover $k$ variables with high probability at a fixed query difficulty $\bard$? and (2) what kind of querying schemes can achieve this limit, especially with computationally-efficient decoding algorithms?
In this paper, we answer these questions for two recovery conditions: exact recovery and almost exact recovery. 
Exact recovery aims to recover all the $k$ variables with high probability as $k\to\infty$. On the other hand, almost exact recovery aims to recover only a fraction $(1-\delta)k$ of variables where $\delta=o(1)$ as $k\to\infty$. We analyze the fundamental trade-offs between recovery accuracy $\delta$, query difficulty $\bard$ and the sample complexity $n$.

\subsection{Main Contributions}
Our main contributions can be summarized as follows.
First, we show that the sample complexity $n$ that is necessary (Proposition~\ref{prop:suff}) and sufficient (Theorem \ref{thm:main}) for exact recovery  scales as
\beq\label{eqn:mainres}
n=c_0\cdot \max\left\{k,(k\log k)/{\bar{d}}\right\}
\eeq
for a constant $c_0>0$. 
We show that the optimal sample complexity for exact recovery is achieved using parity queries, for any query-degree (subset size) distributions when the maximum query degree is bounded above by $\log k$, or for any maximum query degree $D\in\{2,3,\dots,k\}$ when the query-degree distribution follows a particular subset size distribution, defined in~\eqref{eqn:modSol}. 
Note that for $\bar{d}= O(\log k)$, the optimal sample complexity $n$ is inversely proportional to the query difficulty $\bar{d}$. In particular, when the query difficulty is $\bar{d}=\Theta(1)$, the sample complexity scales as $k\log k$, whereas when $\bar{d}=\Theta(\log k)$, the sample complexity scales as $k$.
The achievability results are shown for the  maximum likelihood (ML) decoding rule, which can be implemented by Gaussian elimination algorithm with time steps $O(nk^2)$. 

Second, for almost exact recovery of any fraction $(1-\delta)k$ of variables for $\delta=o(1)$ and $\delta\geq 1/k$, we show that the necessary (Theorem \ref{thm:main2}) and sufficient (Theorem \ref{thm:main3}) sample complexity $n$ scales as
\beq\label{eqn:sam_comp_alex1}
n=c_1\cdot\max\left\{k,{\left(k\log (1/\delta)\right)}/{\bar{d}}\right\}
\eeq for some constant $c_1>0$ independent of $k$, $\bar{d}$ and $\delta$.
In particular we show that this optimal sample complexity for almost exact recovery is achieved with low-complexity belief propagation (BP) decoding rule with time steps $O(n\log k)$. Different from the case of exact recovery, the case of almost exact recovery requires a sample complexity proportional to ${\left(k\log (1/\delta)\right)}/{\bar{d}}$ for $\delta\geq 1/k$ where $\bard\leq \log(1/\delta)$. 
As $\delta$ decreases, the required sample complexity increases logarithmically in $(1/\delta)$. 
In the viewpoint of the classical coupon collecting problem, or the process of throwing balls randomly into bins, it is natural to have the factor $\log (1/\delta)$ instead of $\log k$ in \eqref{eqn:sam_comp_alex1} since heuristically we may stop throwing balls when $(1-\delta)k$ bins are filled; see the case $\bar{d}=1$ in Section \ref{sec:proof3}. 
However, it is highly nontrivial to find a query degree distribution $(\Omega_1,\dots,\Omega_k)$ for which we can control the average query degree $\bard=\sum_{d=1}^k d \Omega_d$, and to show that when the parity queries are generated by this distribution a BP decoding process can successfully decode $(1-\delta)k$ variables at the optimal sample complexity.
The proof idea of this achievability result is an extension of the related result in \cite{luby2002lt}, where it was shown that a BP decoding process can decode all the $k$ variables with $n=k(1+\epsilon)$ parity queries for any $\epsilon>0$ when the parity queries are generated by a soliton distribution for which the query difficulty is fixed as high as $\bard=\Theta(\log k)$. We modify this distribution to control the average query degree so that $\bard=O(\log k)$, possibly $\bard \ll \log k$, and prove that  BP decoder does not stop until we recover $(1-\delta)k$ variables even with the (modified) soliton distribution as in \eqref{eqn:modSol}.


\subsection{Related Works}

\subsubsection{Planted Constraint Satisfaction Problem (CSP)}
The recovery of discrete variables from a random set of measurements is an example of a more general problem, called a planted constraint satisfaction problem (CSP), which has been a subject of intense study in computer science, probability theory, and statistical physics, motivated by clustering, community detection, and cryptography~\cite{abbe2013conditional,achlioptas2008algorithmic,haanpaa2006hard}. In particular, our problem has a connection to a random planted $d$-XOR-satisfiability (XORSAT) problem~\cite{ibrahimi2015set}, which aims to recover $k$ binary variables (the planted solution) satisfying a set of $n$ constraints each of which is that the XOR of size-$d$ subset of variables, chosen uniformly at random among $k\choose d$ possibilities,  should be equal to 0 (or 1).

When $d=2$, this problem has been extensively studied in the context of community detection.
In~\cite{abbe2014decoding}, it was shown that the information-theoretic limit on the optimal sample complexity (the number of parity measurements to recover the $k$ variables) is $n=\frac{1}{2}\frac{k\log k}{(\sqrt{1-\theta}-\theta)^2}$ for $d=2$ where the measurement is corrupted by Bernoulli($\theta$) noise.
In~\cite{mazumdar2017clustering}, the case when the number of communities (sort of labels) can be larger than 2 was considered and the information-theoretic lower bound on the number of pairwise queries ($d=2$) and inference algorithms that closely match this bound were provided. 
The parity queries with query degree $d>2$ has also been considered in the context of locally-encodable coding \cite{mazumdar2017semisupervised} or community recovery in hypergraphs~\cite{dembo2008finite,ahna2019community}. In~\cite{watanabe2013message}, the case of $d=3$ was studied and in~\cite{abbe2013conditional} the phase transition on the optimal sample complexity was demonstrated for an even $d$. In~\cite{ahna2019community}, it was shown that the optimal sample complexity for parity queries with a fixed query degree $d$ scales as $\frac{k\log k}{d}$ as $d$ increases up to order of $\Theta(\log k)$. 

We generalize this problem such that the subset size (query degree) $d$ can be varying over queries. 
This generalization not only models much general querying scenarios including answerers of different abilities but also allows the recovery of binary variables from parity measurements by using a computationally-efficient BP-type algorithm. 
When $d>1$ is fixed over queries, Gaussian elimination, which is the best known inference algorithm to recover $k$ variables from $n$ randomly selected parity measurements of degree-$d$, requires $O(nk^2)$ time steps. Since $n$ should be at least $k$, it requires $O(k^3)$. 
On the other hand, we propose a set of queries (an optimal query degree distribution) with the same average query degree $\bard$ for which the computationally-efficient BP algorithm can almost exactly recover the $k$ variables with high probability at the information-theoretically optimal sample complexity $n$ with time steps $O(n\log k)$.
 This result demonstrates the benefit of mixing different query degrees in reducing the computationally complexity of the recovery algorithms.

\subsubsection{Rateless Codes}

Query design for random erasure with unknown erasure probability has also been widely considered in channel coding problems, especially for reliable internet packet transmissions. Over the internet, packets transmitted from the source are randomly lost before they arrive at the destination, and the communication channel can be modeled as binary erasure channels (BEC) with unknown erasure probabilities. 
Due to the long delay issues regarding the acknowledgement-based TCP/IP protocols, which requires feedbacks from the destinations, Fountain codes~\cite{mackay2005fountain}, also known as forward error-correcting erasure rateless codes, have been the subject of much research for reliable packet transmissions. 
For a given set of $k$ input symbols $(x_1,x_2,\dots, x_k)$, Fountain codes produce a potentially limitless number of parity measurements, which are also called output symbols. By using the well-designed Fountain codes, it is guaranteed that from any set of output symbols of size $k(1+\delta)$ with a small overhead $\delta>0$, the input symbols can be recovered with high probability.  
There have been many attempts to design practical Fountain codes that achieve the reliable recovery of information bits with a small overhead and also with low encoding and decoding complexities. The well-known examples of the practical Fountain codes include LT-codes~\cite{luby2002lt} and Raptor codes~\cite{shokrollahi2006raptor}. 

Our problem has a close relation to the design of rateless codes since specifying a sequence of subsets of variables for parity queries is equivalent to designing codewords.
However, despite the similarities, the Fountain code framework does not account for an answerer's limited capability to answer difficult queries.  
This is where the analogy breaks down, motivating our extension of the Fountain code solution strategy. 
Specifically we define query difficulty as the average size of the subset of variables for parity queries and design the information-theoretically optimal querying schemes (rateless codes) for any given query difficulty $\bard\in[1,k]$.

\subsubsection{Other Types of Queries}

There are other types queries that generate simple function outputs on chosen subsets of variables. One of the common types of binary query other than XOR is the homogeneity query (AND or OR query).
Consider $k$ items with binary attributes. 
The homogeneity query receives the answer 1 if all the items in the chosen subset have the same binary attribute and receives 0 otherwise. The homogeneity-query type has been widely studied in the context of group testing \cite{du2000combinatorial,atia2012boolean,cheraghchi2011group}, especially in detecting a small distinguished subset of items having a given attribute (e.g., defective items). Unlike the parity query, the sample complexity required to recover the $k$ variables does not decrease in general even with the increasing query difficulty (average subset size) $\bard$, since the probability that all the items belong to the same class decreases as the subset size increases. In other words, the answer to the homogeneity query is more biased to ``no" as the subset size increases so that the amount of information each homogeneity query can discover decreases as the subset size increases. More precisely, in~\cite{ahna2019community} it was shown that when each homogeneity query selects $d=\Theta(1)$ items the information-theoretic limit on the number of queries required to recover $k$ variables scales as $\frac{2^{d-2}}{d}{k\log k}$. The parity query with the same query degree $d$ is more efficient than the homogeneity query since the required sample complexity scales as $\frac{k\log k}{d}$.

Another common type of query for a chosen subset of variables is the histogram-type query~\cite{dorfman1943detection,wang2016data}, which measures the exact composition of types (e.g., blood types) for a chosen subset of population. In~\cite{alaoui2019decoding}, information-theoretic bounds on the sample complexity to recover $k$ variables from histogram data were provided when the subset size $d$ of each query is as dense as $d=\Theta(k)$. 
For this case, the sample complexity scales as $\Theta({k}/{\log k})$. This shows that when we can exactly measure the compositions of subsets of variables of size $d=\Theta(k)$, the histogram measurements provide about $\Theta(\log k)$ bits per query, whereas the binary query can maximally provide 1 bit of information so that the minimum sample complexity should be at least $\Theta(k)$ even with the query difficulty of $\bard=\Theta(k)$ . 

\subsection{Paper Organization}
The rest of this paper is organized as follows. In Section \ref{sec:model}, we present the model and formulate the binary classification problem with parity queries for a restricted query difficulty. 
In Section \ref{sec:main}, we establish theory (Proposition \ref{prop:suff}, Theorem \ref{thm:main}--Theorem \ref{thm:main3}) on fundamental trade-offs between sample complexity and query difficulty for exact recovery and for almost exact recovery and also present the optimal querying schemes that achieve these fundamental limits.
In Section \ref{sec:proof}, we outline the proofs of the main results, with technical details presented in the Appendices. In Section \ref{sec:sim}, we present numerical studies that illustrate the tightness of our theoretical results. 
In Section \ref{sec:con}, we discuss possible future research directions.

\subsection{Notations}
We use the notation $\oplus$ for XOR of binary variables, i.e., for $a,b\in\{0,1\}$, $a\oplus b=0$ iff $a=b$ and $a\oplus b=1$ iff $a\neq b$.
We denote by $e_j$ the $k$-dimensional unit vector with its $j$-th element equal to 1.
For a vector $\bx$, $\|\bx\|_1$ denotes the number of $1$'s in the vector $\bx$.
For vectors $\bx$ and $\by$, the inner product between $\bx$ and $\by$ is denoted by $\bx \cdot \by$.
For two integers $\beta$ and $\gamma$, we use the notation $\beta\equiv \gamma$ to indicate that $\mod(\beta,2)=\mod(\gamma,2)$.
For two vectors $\bx=(x_1,x_2,\dots, x_k)$ and $\by=(y_1,y_2,\dots, y_k)$,  $\bx\equiv \by$ means that $\mod(x_i,2)=\mod(y_i,2)$ for all $i\in\{1,2,\dots,k\}$.
We use the $o(\cdot)$, $\omega(\cdot)$, $O(\cdot)$, $\Omega(\cdot)$ and $\Theta(\cdot)$ notations to describe the asymptotics of real sequences $\{a_k\}$ and $\{b_k\}$: $a_k=o(b_k)$ is equivalent to the condition that for $\forall M>0$ there exists $k_0$ such that $|a_k|\leq M b_k$ for $\forall k\geq k_0$; 
$a_k=\omega(b_k)$ is equivalent to the condition that for $\forall M>0$ there exists $k_0$ such that $|a_k|> M |b_k|$ for $\forall k\geq k_0$;
$a_k=O(b_k)$ implies that there exists $M>0$ and $k_0$ such that $|a_k|\leq M b_k$ for all $k\geq k_0$; $a_k=\Omega(b_k)$ implies that there exists $M>0$ and $k_0$ such that $|a_k|\geq M b_k$ for all $k\geq k_0$; $a_k=\Theta(b_k)$ implies that there exist $M,M'>0$ and $k_0$ such that $M' b_k\leq a_k\leq M b_k$ for all $k\geq k_0$.
The logarithmic function $\log$ is with base $e$.

\section{Model: Parity Query for Binary Classification}
\label{sec:model}

\begin{figure}[t]
\centerline{\includegraphics[scale=0.5]{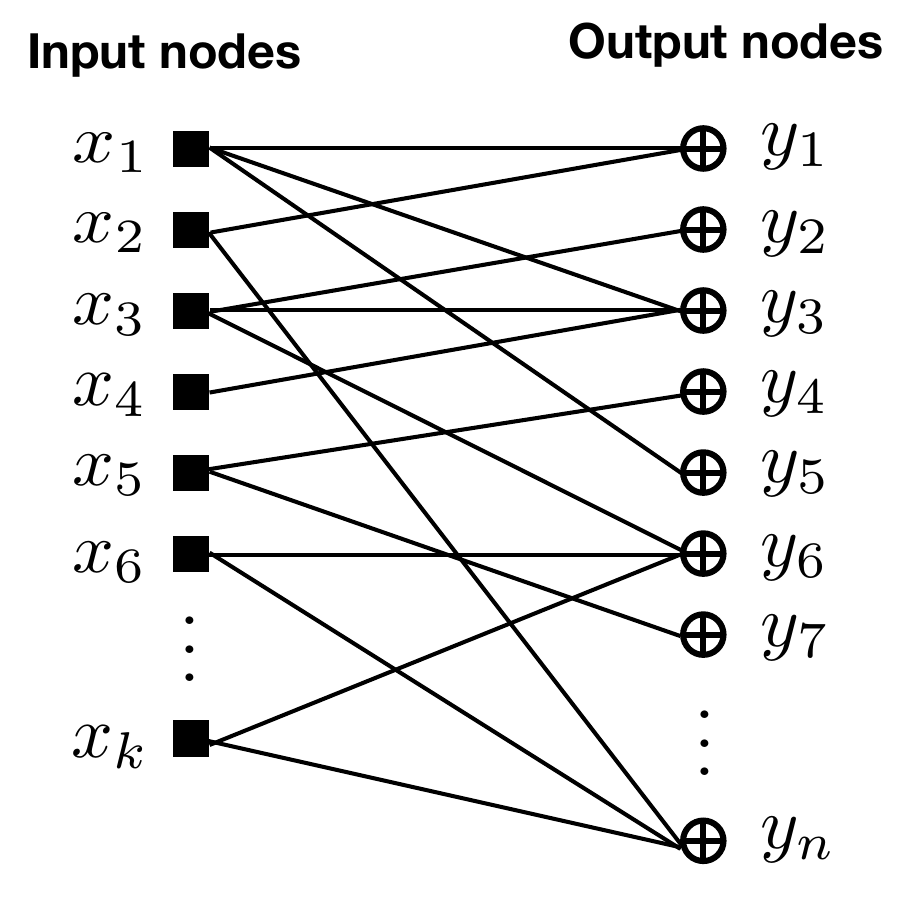}}
\caption{Bipartite graph between input nodes and output nodes, where the input nodes represent the $k$ binary variables that we aim to recover and the $n$ output nodes represent the parity measurements of those variables. The edges between input nodes and output nodes specify the subset of input nodes that generates the parity measurement at each output node.}
\label{fig:graph}
\end{figure}

Consider a $k$-dimensional binary vector $\bx=(x_1,x_2,\dots, x_k)\in\{0,1\}^k$.
We aim to recover the values of $\bx$ by designing parity queries and receiving answers to only a randomly selected subset of the designed queries.
Each parity query is designed independently by first sampling a query degree $d$ from a probability distribution $(\Omega_1,\dots, \Omega_k)$,  where $\Omega_d$ is the probability that the value $d$ is chosen, and then selecting $d$ components of $\bx$ uniformly at random among ${k\choose d}$ possibilities. Each query then asks the parity bit (XOR) of the chosen $d$ elements of $\bx$. The average subset size of parity queries
\beq\label{eqn:qeury_diff}
\bar{d}=\sum_{d=1}^k d\cdot \Omega_d
\eeq
is called {\it query difficulty}. When the answered queries are denoted by  $\by=(y_1,y_2,\dots, y_n)\in\{0,1\}^n$,  the relation between the $k$ binary variables and $n$ answered queries can be depicted by a bipartite graph with $k$ input nodes on one side and $n$ output nodes on the other side as shown in Fig.~\ref{fig:graph}. The query difficulty $\bard$ is the average degree (the number of edges) of output nodes.

The process of recovering  $\bx$ from the collected measurements $\by$ is called inference or decoding. 
Let $\hat{\mathbf{x}}(\by)=(\hat{x}_1(\by),\hat{x}_2(\by),\dots, \hat{x}_k(\by))$ denote the estimate of $\bx$ given $\by$ and define  the agreement measure $A(\bx, \hat{\mathbf{x}}(\by))$ between $\bx$ and $\hat{\bx}(\by)$ by the fraction of common components between $\bx$ and $\hat{\bx}(\by)$, i.e.,
\beq
A(\bx, \hat{\mathbf{x}}(\by))=\frac{1}{k}\sum_{i=1}^k \mathbb{1}(x_i=\hat{x}_i(\by)).
\eeq
The two recovery accuracies we consider in this paper are defined as below.
\begin{itemize}
\item Exact recovery: Estimator $\hat{\bx}$ exactly recovers $\bx$ if $\Pr(A(\bx, \hat{\mathbf{x}}(\by))=1)\to 1$ as $k\to\infty$.
\item Almost exact recovery of $\alpha$ fraction: For a given $\alpha=1-\delta$ with $\delta=o(1)$, estimator $\hat{\bx}$ almost exactly recovers $\bx$ for $\alpha$ fraction if  $\Pr(A(\bx, \hat{\mathbf{x}}(\by))\geq \alpha)\to 1$ as $k\to\infty$.
\end{itemize}
Define the probability of error for exact recovery and that for almost exact recovery for $\alpha$ fraction as 
\begin{align}
P_e^{(k)}&=\min_{\hat{\bx}(\cdot)}\Pr(A(\bx, \hat{\mathbf{x}}(\by))\neq 1),\label{eqn:err_prob}\\
P_{e,\alpha}^{(k)}&=\min_{\hat{\bx}(\cdot)}\Pr(A(\bx, \hat{\mathbf{x}}(\by))< \alpha),\label{eqn:alerr_prob}
\end{align}
respectively.
The minimum number $n$ of measurements required to guarantee $P_e^{(k)}\to 0$  as $k\to\infty$, minimized over all output degree distributions $(\Omega_1,\dots, \Omega_k)$ for a fixed $k$ and $\bard$, is called \emph{sample complexity for exact recovery}, and that for $P_{e,\alpha}^{(k)}\to 0$ is called \emph{sample complexity for almost exact recovery of $\alpha$ fraction}.
In Section \ref{sec:main}, we establish fundamental limits on the sample complexities for the two recovery conditions for any fixed value of query difficulty $\bar{d}\in[1,k]$.

\section{Main Results}
\label{sec:main}


\subsection{Exact Recovery}
In this section,  we establish necessary and sufficient conditions on the sample complexity $n$ guaranteeing exact recovery for a fixed query difficulty $\bard$ in~\eqref{eqn:qeury_diff}. 

We first state the necessary condition on the sample complexity $n$. This condition holds as long as every query is generated independently from an identical distribution.

\begin{proposition}\label{prop:suff}
{\it 
To reliably recover binary vector $\bx\in\{0,1\}^k$ with $P_e^{(k)}\leq 1/k^u$ for some constant $u>0$ by using i.i.d. parity queries  with average query difficulty $\bard$, it is necessary that
\beq\label{eqn:lower}
n\geq c_l\cdot \max\left\{k,\frac{k\log k}{\bar{d}}\right\},
\eeq
for some constant $c_l>0$, independent of $k$ and $\bar{d}$. 
}
\end{proposition}
The proof of Proposition \ref{prop:suff} will be presented in Section \ref{sec:proof0}. 

Showing the necessity of $n\geq k$ is straightforward. Each parity query $y_i$ represents a linear equation (with coefficients in $\mathbb{F}_2$) that the $k$ unknown binary variables $(x_1,\dots, x_k)$ should satisfy. Since there are $k$ unknowns, it is necessary to have at least $n=k$ linear constraints to have a unique solution.  The necessity of $n\geq (c_l {k\log k})/{\bard}$ follows from a property of random graphs. In the bipartite graph between input nodes and output nodes as in Fig.~ \ref{fig:graph}, we say that an input node is isolated if it is not connected to any of the output nodes. 
The error probability $P_e^{(k)}$ is bounded below by 1/2 times the probability that a fixed input node is isolated, since when an input node is isolated the exact recovery of $\bx$ is impossible with at least 1/2 probability. We find a probability that an input node is isolated  when the average query degree of output nodes is $\bard$, and derive the necessary condition on $n$ to make  this probability smaller than $1/k^u$ for some constant $u>0$  as $k\to\infty$.


One of the main contributions of this paper is to show that the bound in~\eqref{eqn:lower} is indeed achievable (up to constant scaling) for any $\bar{d}$ from $\Theta(1)$ to $\Theta(\log k)$ by properly designed parity querying strategies. 
We state the achievability results in two different settings, depending on whether or not there exists a constraint on the maximum query degree.
In the first setting, we assume that the maximum degree $D$ of each query is bounded above by $\log k$.
For this case, we show that the exact recovery is guaranteed with the sample complexity~\eqref{eqn:lower} (with a different constant scaling) for any degree distribution $(\Omega_1,\Omega_2,\dots, \Omega_D)$ with $\bard=\sum_{d=1}^D d\Omega_d\leq \log k$. 

When the maximum query degree is not smaller than $\log k$, on the other hand, there exists a simple counterexample that shows that not all $(\Omega_1,\Omega_2,\dots, \Omega_k)$ can guarantee the exact recovery with the sample complexity~\eqref{eqn:lower}.  
For example, consider the case when there are $n\gamma$ degree-1 queries and $n(1-\gamma)$ degree-$k$ queries. The degree-$k$ queries do not provide any new information other than $\mod(\sum_{i=1}^k x_i,2)$, and thus to recover all the $k$ variables by degree-1 queries, which asks one of the randomly picked variable at each querying, it is required to collect $n\gamma=\Theta(k\log k)$ answers as shown in~Proposition \ref{prop:suff}. For $\gamma=\frac{k-\log k}{k-1}$, the average query degree $\bard$ for this querying scheme is $\log k$, but still $n=\Theta(k\log k)$ answers are required, and thus the bound~\eqref{eqn:lower} is not achievable by this degree distribution. For this second case where the maximum query degree can be any number $D\in\{2,3,\dots, k\}$, we provide a particular output degree distribution $(\Omega_1,\dots, \Omega_k)$ for which we can control the query difficulty $\bar{d}$ from $\Theta(1)$ to $\Theta(\log k)$ and show that the exact recovery is guaranteed with sample complexity,
$
n= c_u\cdot \max\left\{k,\frac{k\log k}{\bar{d}}\right\},
$
for some constant $c_u>0$, possibly different from $c_l$ in~\eqref{eqn:lower}.


Suppose that distribution $(\Omega_1, \ldots, \Omega_k)$  is set to the soliton distribution with the maximum degree $D$
\beq\label{eqn:modSol}
\begin{aligned}
\Omega_d = 
	\begin{cases}
	\frac{1}{D} & \text{ if } d=1 \\
	\frac{1}{d(d-1)} & \text{ if } 2\leq d \leq D \\
	0 & \text{ if } d > D,
	\end{cases}
\end{aligned}
\eeq
for some $D\in\{2,3,\dots,k\}$. Here, for simplicity, we assume that $k\geq 3$.
Note that under the soliton distribution  the query difficulty scales as $\log D$ since
\beq\label{eqn:bard_loup}
\log (D+1) < \bar d = \frac{1}{D} + \sum_{d=2}^{D} \frac{1}{d-1} = \sum_{d=1}^D \frac{1}{d} < \log D + 1.
\eeq
Therefore, as $D$ increases from $2$ to $k$, the query difficulty $\bard$ scales from $\log 3$ to $\log k$.

\begin{theorem}\label{thm:main}
{\it
Assume that the parity-based querying strategy chooses $d$ variables uniformly at random, where for each query the query degree $d$ is randomly selected according to the distribution $(\Omega_1,\dots,\Omega_D)$. When the maximum query degree $D\leq \log k$, the exact recovery is guaranteed, i.e., $P_e^{(k)}\to0$ as $k\to\infty$, by maximum likelihood (ML) decoding rule for any degree distribution $(\Omega_1,\dots,\Omega_D)$ satisfying $\bard=\sum_{d=1}^D d\Omega_d\leq \log k$ when sample complexity satisfies
\beq\label{eqn:sam_up_max}
n\geq (5\log 2)(1+\epsilon)\frac{k\log k}{\bard}
\eeq
for a small constant $\epsilon>0$, independent of $k$ and $\bard$. 

Moreover, when the query degree $d$ is sampled by the soliton distribution~\eqref{eqn:modSol}, for any maximum query degree $D\in\{2,\dots, k\}$ the exact recovery is guaranteed by ML decoding rule when sample complexity
\beq\label{eqn:sam_up}
n=c_u \cdot \max\left\{k,\frac{k\log k}{\bar{d}}\right\}
\eeq
for some universal constant $c_u>0$, independent of $k$ and $\bard$. 
}
\end{theorem}
The proof of~Theorem \ref{thm:main} will be presented in Section \ref{sec:proof1}.

Combined with Proposition \ref{prop:suff}, Theorem \ref{thm:main} shows that the number $n$ of measurements in~\eqref{eqn:sam_up} is optimal up to constants for exact recovery of $k$ variables for a fixed query difficulty $\bard$.
This establishes the optimality of the specified soliton-distributed random subset selection rule for all $\bard=\Theta(1)$ to $\Theta(\log k)$. 
When the query difficulty is $\bard = O(\log k)$, the sample complexity $n$ to reliably recover $k$ binary variables is inversely proportional to the query difficulty $\bard$. 
When the query difficulty does not increase with $k$, i.e., $\bard=\Theta(1)$, it is necessary and sufficient to have $n=\Theta(k\log k)$.
In this regime, the ratio between $k$ and $n$ converges to 0 as $k\to\infty$.
On the other hand, when we increase the query difficulty to $\bard=\Theta(\log k)$, it is enough to have $n=\Theta(k)$, which results in a positive limit of $k/n$ as $k\to\infty$.


The ML decoding rule assumed in proving Theorem \ref{thm:main} can be implemented by using Gaussian elimination, which solves an inverse problem that recovers $\bx\in\{0,1\}^k$ by using $\by\in\{0,1\}^n$ where each $y_i$ represents a linear equation of $\bx$ with coefficients in $\mathbb{F}_2$. The time complexity of Gaussian elimination is $O(nk^2)$. 
Of interest is whether we can further reduce this time complexity for a properly designed query degree distribution and a decoding algorithm.
In the next section, we show that when the degree distribution follows the soliton distribution \eqref{eqn:modSol}, BP decoding algorithm with time complexity  $O(n\log k)$ guarantees the almost exact recovery of $\alpha=(1-\delta)$ fraction of total variables for any $\delta=o(1)$ at the optimal sample complexity.

\subsection{Almost Exact Recovery}\label{sec:IIIB}

The aim of almost exact recovery is to reliably recover a fraction $\alpha=(1-\delta)$ of total variables where $\delta=o(1)$ as $k\to\infty$ and $\delta\geq 1/k$. 
We first find necessary conditions on $n$ to guarantee $P_{e,\alpha}^{(k)}\to 0$ as $k\to\infty$ for a fixed query difficulty $\bard\in[1,k]$.
\begin{theorem}\label{thm:main2}
{\it 
Assume that the parity-based query design strategy chooses $d$ variables uniformly at random, where for each query the query degree $d$ is randomly selected according to a distribution $(\Omega_1,\dots,\Omega_k)$ satisfying $\bard=\sum_{d=1}^k d\Omega_d$ and $\sum_d d^2 \Omega_d \ll k \bard$. Then to reliably recover a fixed proportion $\alpha k =(1-\delta)k$ of variables, i.e., for $P_{e,\alpha}^{(k)}\to0$ as $k\to\infty$ for $\delta=o(1)$,  the sample complexity $n$ must satisfy
\beq\label{eqn:lower_al}
n\geq(1-\epsilon)\cdot\max\left\{k(1-H_{\sf B}(\delta)-\delta),\frac{k\log (1/\delta) }{\bar{d}}\right\},
\eeq
for any $\epsilon>0$ independent of $k,\delta$ and $\bard$, where $H_{\sf B}(\delta)=-\delta\log\delta-(1-\delta)\log(1-\delta)$.
}
\end{theorem}
The proof of~Theorem \ref{thm:main2} will be presented in Section \ref{sec:proof2}.

Theorem \ref{thm:main2} provides two kinds of necessary conditions on the sample complexity $n$. The first necessary condition, $n\geq (1-\epsilon)k(1-H_{\sf B}(\delta)-\delta)$, is from Fano's inequality~\cite{cover2012elements}, and this condition does not depend on the query difficulty $\bard$.
The second necessary condition, $n\geq (1-\epsilon) \frac{k\log (1/\delta) }{\bar{d}}$, follows from  the fact that $P_{e,\alpha}^{(k)}$ can never tend to 0 if more than $\delta k$ input nodes are isolated, i.e., not connected to any of $n$ output nodes of average degree $\bard$, in the random bipartitie graph. 
Compared to the necessary condition for exact recovery (Proposition \ref{prop:suff}), the sample complexity for almost exact recovery requires $n\geq k\log(1/\delta)/\bard$ instead of $n\geq c_l k\log k/\bard$ for $\delta\geq 1/k$.

We next consider sufficient conditions on $n$ to guarantee $P_{e,\alpha}^{(k)}\to 0$. We show that when the degree of the output nodes is generated by the soliton distribution~\eqref{eqn:modSol}, almost exact recovery of $\alpha=1-\delta$ fraction can be achieved with the sample complexity~\eqref{eqn:sam_up2} via using a belief-propagation (BP) decoder. The BP decoder is computationally much more efficient than the Gaussian elimination decoder (ML decoder) and the required time steps reduces from $O(nk^2)$ to $O(n\log k)$. 
 
Referring to the bipartite graph between input nodes and output nodes shown in Fig.~\ref{fig:graph}, the BP decoding process works as follows. 
We use terminologies from~\cite{luby2002lt} where the BP decoding for LT codes was analyzed. 
An input node is said to be uncovered if its value is unknown, and to be covered if its value is known. All input nodes are initially uncovered. 
At the first step the BP decoder finds all output nodes of degree one and recovers the values of their unique neighboring input nodes. 
At each subsequent step the decoder {\emph{processes}} one covered input node by adding (XORing) its value to all its connected output nodes and then removing it (with all its edges) from the graph. We say that an output node is {\emph{released}} by the processing of the input node if its degree was larger than 1 before the processing, and it is equal to 1 after the processing. The newly released output nodes may increase the number of the covered input nodes if their unique neighboring input nodes have not yet been covered. If the number of covered input nodes that have not yet been processed does not drop to 0 until $\alpha k$ input nodes are processed, the decoding of $\alpha k$ input nodes is successful. 
In the following theorem, we prove sufficient conditions on the sample complexity $n$ for this BP decoding algorithm to be successful for almost exact recovery of $\alpha$ fraction, where the output nodes are generated independently from the query degree distribution following a soliton distribution~\eqref{eqn:modSol}.

\begin{theorem}\label{thm:main3} 
{\it
Assume that the parity-based query design chooses $d$ variables uniformly at random, where for each query the query degree $d$ is randomly selected according to the soliton distribution~\eqref{eqn:modSol}. Then, if the average query difficulty satisfies $\bard>1$, at least $\alpha k=(1-\delta) k$ variables can be reliably recovered via using the BP decoder, i.e., $P_{e,\alpha}^{(k)}\to0$ as $k\to\infty$ for $\delta=o(1)$, with sample complexity
\beq\label{eqn:sam_up2}
n= c\cdot \max\left\{k,\frac{k\log (1/\delta)}{\bard-1}\right\},
\eeq
for some constant $c>0$, independent of $k,\delta,$ and $\bard$. If $\bard=1$, it is sufficient to have $n=(1+\epsilon)k\log(1/\delta)$ for any $\epsilon>0$. 
}
\end{theorem}
The proof of~Theorem \ref{thm:main3} will be presented in Section \ref{sec:proof3}.

\vspace{0.1in}
\begin{remark}
The main difference in the required sample complexities between exact recovery (Theorem \ref{thm:main}) and almost exact recovery (Theorem \ref{thm:main3}) is in the second factors in~\eqref{eqn:sam_up} and~\eqref{eqn:sam_up2}, i.e., for exact recovery the sample complexity should be a constant scaling of  $(k\log k)/\bard$ but for almost exact recovery the sample complexity should be a constant scaling of $(k\log(1/\delta))/\bard$. We have $\delta=o(1)$ and $\delta\geq \frac{1}{k}$ since $k\delta$ should be at least 1. 
As $\delta$ decreases, the required sample complexity for almost exact recovery of $(1-\delta)$ fraction increases logarithmically in $(1/\delta)$, and the required sample complexity becomes the same as that of exact recovery when $\delta=1/k$.
\end{remark}
\vspace{0.1in}

Theorem \ref{thm:main2} and Theorem \ref{thm:main3} state that to guarantee reliable recovery of a fraction $\alpha=(1-\delta)$ of total variables it is necessary and sufficient that the sample complexity be $n= c_1 \cdot \max\left\{k,\frac{k\log (1/\delta)}{\bar{d}}\right\}$ for some constant $c_1>0$ where the query difficulty is $\bard$.
Again, we can observe a fundamental trade-off between the query difficulty and the sample complexity. 
For $\bard=O(\log (1/\delta))$, the sample complexity $n$ for almost exact recovery is inversely proportional to the query difficulty $\bard$.

The main technical difficulty in the proof of Theorem \ref{thm:main3} lies in the control of $r(L)$, the probability that an output is released when $L$ input nodes remain unprocessed (see \eqref{eqn:defrL}); if $r(L)$ is not sufficiently large, the decoding process may stop before recovering $\alpha$ fraction. In \cite{luby2002lt}, the lower bound for $r(L)$ was obtained with ease since the degree distribution is given by a combination of the robust soliton distribution and the ideal soliton distribution, where the ideal part guarantees that $r(L)$ is bounded below by $1/k$. However, for the almost exact recovery, we cannot follow this approach since $\bard=\Omega(\log k)$ when the ideal soliton distribution is included in the degree distribution, while the query difficulty we consider can be $\bard\ll \log k$. Thus, we need to prove a lower bound for $r(L)$ without using the ideal soliton distribution but with the modified soliton distribution in \eqref{eqn:modSol}. See Lemma \ref{lem:rL} and Appendix~\ref{app:lem:rL} for the lower bound and the proof for it.

\section{Proof of Main Results}
\label{sec:proof}

\subsection{Proof of Proposition \ref{prop:suff}: Necessary Conditions for Exact Recovery}\label{sec:proof0}


We prove the necessary condition $n\geq (c_l {k\log k})/{\bard}$  by using the fact that the error probability $P_e^{(k)}$ is bounded below by 1/2 times the probability that a fixed input node is isolated.
In~\cite{shokrollahi2006raptor}, the probability that a fixed input node is isolated was calculated under the assumption that the edges from each output node are connected to input nodes uniformly at random. 
 We first review this special case for the purpose of generalizing~\cite{shokrollahi2006raptor} to arbitrary distributions. 
Consider a fixed input node and an output node of degree $d$. The probability that the fixed input node is not connected to this output node of degree $d$ equals $1-d/k$ where $d$ input nodes are chosen uniformly at random among the total $k$ input nodes. Since an output node has degree $d$ with probability $\Omega_d$, the probability that the fixed input node is not connected to an output node equals
\beq
\sum_{d=0}^k \Omega_d (1-d/k)=1-\bar{d}/k.
\eeq
Since there are $n$ output nodes and these output nodes are sampled independently, the probability that the fixed input node is isolated (not connected to any of those output nodes) equals
\beq\label{eqn:iso_pro}
\left(1-\frac{\bar{d}}{k}\right)^{n}.
\eeq
By the mean value theorem, it can be shown that $(1-\bar{d}/{k})^{n}\geq e^{-\beta/(1-\beta/n)}$ where $\beta=n\bar{d}/k$.
To satisfy $P_e^{(k)}\leq 1/k^u$ for some constant $u>0$, it is necessary that $e^{-\beta/(1-\beta/n)}/2\leq 1/k^u$, which is equivalent to
\beq
\begin{split}
\beta&\geq \log k\cdot\frac{u-\log2/\log k}{1+(u\log k-\log 2)/n}\\
&\geq c_l\log k,
\end{split}
\eeq
for some constant $c_l>0$.
By plugging in $\beta=n\bar{d}/k$, we get
\beq\label{eqn:nec_LT_nbard}
n\bar{d}\geq c_l k\log k.
\eeq

We turn to the more general case where the $d$ edges from an output node are connected to $d$ input nodes chosen by non-uniform random selection. Consider an input node that has the smallest probability (among all the $k$ input nodes) to be connected to an output node. For such an input node, the probability that this input node is isolated is larger than the probability in~\eqref{eqn:iso_pro}, which was derived under the assumption of uniform random selection.
Since there should be no isolated node to guarantee the exact recovery with high probability, the condition~\eqref{eqn:nec_LT_nbard} on the sample complexity is necessary not only for uniform random selection but also for any distribution as long as the output nodes (the parity queries) are generated independently by an identical distribution with average query difficulty $\bard$.

\subsection{Proof of~Theorem \ref{thm:main}: Sufficient Conditions for Exact Recovery}\label{sec:proof1}
In this section, we prove~Theorem \ref{thm:main} by providing an upper bound on $P_e^{(k)}$ and  showing that the sample complexity $n$ sufficient to make this upper bound converge to $0$ as $k\to\infty$ is equal to \eqref{eqn:sam_up_max} when the maximum query degree $D\leq \log k$ and is equal to \eqref{eqn:sam_up}  when the query degree distribution follows the soliton distribution \eqref{eqn:modSol} for any maximum query degree $D\in\{2,3,\dots,k\}$.
 
Consider $P_e^{(k)}$ defined in~\eqref{eqn:err_prob}. We consider the maximum likelihood (ML) decoding rule for $\hat{\bx}(\cdot)$.
Assume that we collect $n$ output symbols $(y_1,\dots, y_n)$ each of which equals $y_i=\mod\left(\sum_{j=1}^n v_{ij} x_j,2\right)$ where $v_{ij}=1$ if the $j$-th item is included in the $i$-th query and $v_{ij}=0$ otherwise. Consider a matrix $A$ whose $i$-th row is $\bv_i=(v_{i1},v_{i2},\dots, v_{ik})$, i.e.,
\beq
A:=[\bv_1;\bv_2;\dots; \bv_n].
\eeq
We call $A$ a sampling matrix. Given $(y_1,\dots, y_n)$ and the sampling matrix $A$, the ML decoding rule finds $\bx=(x_1,x_2,\dots, x_k)^T\in\{0,1\}^k$ such that
\beq
A\bx\equiv (y_1,y_2,\dots,y_n)^T.
\eeq
If there is a unique solution $\bx\in\{0,1\}^k$ for this linear system, then we set $\hat{\bx}(\by)=\bx$. If there is more than one $\bx$ satisfying this linear system, then an error is declared. 
The probability of error is thus equal to
\beq
P_e^{(k)}=\sum_{\bx\in\{0,1\}^k} \Pr(\bx)\Pr(\exists \bx' \neq \bx \text{ such that }A\bx'\equiv A\bx).
\eeq
Due to symmetry of the sampling matrix $A$, the probabilities $\Pr(\exists \bx' \neq \bx \text{ such that }A\bx'\equiv A\bx)$ are equal for every $\bx\in\{0,1\}^k$. Thus, we focus on the case where $\bx$ is the vector of all zeros and consider
\beq
P_e^{(k)}=\Pr(\exists \bx' \neq \bzero \text{ such that }A\bx'\equiv \bzero).
\eeq
By using the union bound, it can be shown that
\beq
\begin{aligned}
P_e^{(k)}
&\leq \sum_{\bx'\neq \bzero} \Pr(A\bx'\equiv \bzero)=\sum_{s=1}^k \sum_{\|\bx'\|_1=s}\Pr( A\bx'\equiv \bzero )\\
&=\sum_{s=1}^k { k \choose s} \Pr\left(A\left(\sum_{i=1}^s e_i\right) \equiv \bzero\right)
\end{aligned}
\eeq
where $e_i$ is the $i$-{th} standard unit vector. The last equality follows from the symmetry of the sampling matrix $A$. 
Since all the parity queries are generated independently by the identically distributed $\bv_i$'s, each of which has weight $d$ with probability $\Omega_d$,
\beq
\begin{aligned}\label{eqn:P_e_m1}
P_e^{(k)}&\leq \sum_{s=1}^k { k \choose s}  \left(  \Pr\left( \bv_1\cdot \left(\sum_{i=1}^s e_i\right) \equiv 0\right) \right)^n\\
&=\sum_{s=1}^k { k \choose s}
\left(\sum_{d=1}^k \Omega_d \Pr\left(\bv_1\cdot \left(\sum_{i=1}^s e_i\right) \equiv 0\Bigg| \|\bv_1\|_1=d\right) \right)^n.
\end{aligned}
\eeq
We next analyze 
\beq
\Pr\left(\bv_1\cdot \left(\sum_{i=1}^s e_i\right)\equiv0 \Bigg| \|\bv_1\|_1=d\right).
\eeq
Note that $\bv_1\cdot \left(\sum_{i=1}^s e_i\right)\equiv0$ if and only if there are even number of 1's in the first $s$ entries of $\bv_1$. When the $d$ non-zero positions of $\bv_1$ is selected uniformly at random among $k\choose d$ possibilities, this probability equals
\beq
\begin{split}\label{eqn:single}
&\Pr\left(\bv_1\cdot \left(\sum_{i=1}^s e_i\right)\equiv 0 \Bigg| \|\bv_1\|_1=d\right)=\frac{\sum_{\substack{i\leq d\\ i \text{ is even}}} {s \choose i}{k-s \choose d-i} }{{k \choose d}}.
\end{split}
\eeq

We next provide an upper bound on~\eqref{eqn:single}. Define
\beq
I_d = \sum_{\substack{i\leq d\\ i \text{ is even}}}{s\choose i}{k-s\choose d-i}.
\eeq
In the following lemma, we provide an upper bound on $I_d$ as a multiple of ${k\choose d}$. The proof of this lemma is based on that of a similar lemma provided in~\cite{ahna2019community}, where the upper bound on $I_d$ is stated depending on the regimes of $s$ for a fixed $d$. We provide an alternative version where the upper bound on $I_d$ depends on the regimes of $d$ for a fixed $s$. 

\begin{lemma}\label{lem:comb}
{\it
Consider the case that $s \leq \frac{k}{2}$ (i.e., $s \leq k-s$). Define 
\beq\kappa(s) = \frac{k-s+1}{2s+1}.\eeq
\begin{enumerate}
\item For $d \leq \frac{k}{2}$ (or, $k-d \geq d$), when we define $\beta = \frac{k-d+1}{d},$
	\beq
	I_d \leq \begin{cases}
	\left(1- \frac{2s}{5\beta}\right){k\choose d},& \text{when } d < \kappa(s), \\
	\frac{4}{5}{k\choose d}, & \text{when } d \geq \kappa(s).
	\end{cases}
	\eeq
\item For $d > \frac{k}{2}$ (or, $k-d < d$), when we define  $\beta' = \frac{d+1}{k-d}$,
	\beq
	I_d \leq \begin{cases}
	\left(1- \frac{2s}{5\beta'}\right){k\choose d},& \text{when } d > k-\kappa(s), \\
	\frac{4}{5}{k\choose d}, & \text{when } d \leq k-\kappa(s).
	\end{cases}
	\eeq
\end{enumerate}
In the case $s > \frac{k}{2}$, we can obtain the bounds for $I_d$ simply by changing $s$ to $k-s$.
}
\end{lemma}
\begin{IEEEproof}
Appendix~\ref{app:lem:comb}.
\end{IEEEproof}

By using Lemma \ref{lem:comb} and~\eqref{eqn:single}, the upper bound on $P_e^{(k)}$ in~\eqref{eqn:P_e_m1} can be further bounded by
\beq \begin{split} \label{eqn:Sigma_s bound}
P_e^{(k)} &\leq 2\sum_{s \leq \frac{k}{2}} { k \choose s}\left(\sum_{d=1}^{\ceil*{\kappa(s)}-1} \left(1- \frac{2s}{5\beta}\right) \Omega_d+ \sum_{d=\ceil*{\kappa(s)}}^{k-\ceil*{\kappa(s)}} \frac{4}{5}\Omega_d
+ \sum_{d=k-\ceil*{\kappa(s)}+1}^{k} \left(1- \frac{2s}{5\beta'}\right) \Omega_d \right)^n \\
&= 2\sum_{s \leq \frac{k}{2}} { k \choose s} (1-\Sigma_s)^n \leq 2\sum_{s \leq \frac{k}{2}} { k \choose s} e^{-n \Sigma_s},
\end{split} \eeq
where we let
\beq\label{eqn:sig_s}
\begin{split}
\Sigma_s &= \frac{1}{5} \sum_{d=\ceil*{\kappa(s)}}^{k-\ceil*{\kappa(s)}} \Omega_d + 
\frac{2s}{5} \left( \sum_{d=1}^{\ceil*{\kappa(s)}-1} \frac{d\,\Omega_d}{k-d+1} + \sum_{d=k-\ceil*{\kappa(s)}+1}^{k} \frac{(k-d)\Omega_d}{d+1} \right).
\end{split}
\eeq

We next find the sufficient conditions on the sample complexity $n$  for two different cases: 1) for any query degree distribution with the maximum query degree $D\leq \log k$ and 2) for the soliton distribution defined in~\eqref{eqn:modSol} for any maximum query degree $D\in\{2,3,\dots, k\}$.

{\bf Case 1:} For any query degree distribution, when the maximum query degree $D\leq c_d\log k$ for some $0<c_d\leq 1$ we can prove the following lemma. 
\begin{lemma}\label{lem:sam_case1} With the sample complexity
\beq\label{eqn:sam_case1}
n\geq (5\log 2)(1+\epsilon)\frac{k\log k}{\bard}
\eeq
for a small universal constant $\epsilon>0$, the term ${k \choose s}e^{-n\Sigma_s}$ is bounded above as follows. 
\begin{enumerate}
\item When $1\leq s\leq \lfloor \frac{k}{2c_d \log k}\rfloor$,
\beq
\binom{k}{s} e^{-n\Sigma_s} <  (1/k^\epsilon)^s.
\eeq
\item When $\lfloor \frac{k}{2c_d \log k}\rfloor+1\leq s\leq k/2$,
\beq
\binom{k}{s} e^{-n\Sigma_s} <  \exp(-\epsilon(\log 2)k).
\eeq
\end{enumerate}
\end{lemma}
\begin{IEEEproof}
Appendix~\ref{app:lem:sam_case1}.
\end{IEEEproof}
From Lemma \ref{lem:sam_case1}, when the sample complexity $n$ satisfies \eqref{eqn:sam_case1} we can further bound $P_e^{(k)}$ in~\eqref{eqn:Sigma_s bound} by
\beq
P_e^{(k)} \leq 2\sum_{s=1}^{\lfloor \frac{k}{2c_d\log k}\rfloor} \left(\frac{1}{k^\epsilon}\right)^s+2 \sum_{s=\lfloor \frac{k}{2c_d\log k}\rfloor+1}^{\frac{k}{2}}\exp\left(-\epsilon (\log 2)k\right)\leq \frac{2}{k^\epsilon-1}+k\exp(-\epsilon (\log 2)k),
\eeq
which goes to 0 as $k\to \infty$.

{\bf Case 2:} Suppose that distribution of the query degree $d$ is specified by a soliton distribution defined in~\eqref{eqn:modSol} for some $D\in\{2,3,\dots,k\}$.
Here, we assume that $k\geq 3$.
For this soliton distribution, we provide an upper bound on ${k \choose s}e^{-n\Sigma_s}$ in~\eqref{eqn:Sigma_s bound} for $s \leq \frac{k}{2}$ depending on the regime of $\ceil*{\kappa(s)}$ with conditions on the sample complexity $n$.
\begin{lemma}\label{lem:kappas}
{\it
With the sample complexity
\beq\label{eqn:sam_suff1}
n\geq c_u\max\left\{k,\frac{k\log k}{\bar{d}}\right\}
\eeq
for some constant $c_u>0$, the term ${k \choose s}e^{-n\Sigma_s}$ is bounded above as follows. 
\begin{enumerate}
\item If $\ceil*{\kappa(s)} > D$,
\beq
\binom{k}{s} e^{-n\Sigma_s} <  k^{-s}.
\eeq
\item If $4 \leq \ceil*{\kappa(s)} \leq D$
\beq
\binom{k}{s} e^{-n\Sigma_s}  \leq
	\begin{cases}
	k^{-s} & \text{ if } s \leq \sqrt k \,, \\
	2^{-2\sqrt k} &\text{ if } \sqrt k < s \leq k/2 \,.
	\end{cases}
\eeq
\item If $\ceil*{\kappa(s)} \leq 3$,
\beq
\binom{k}{s} e^{-n\Sigma_s} \leq 2^k e^{-k}.
\eeq
\end{enumerate}
}
\end{lemma}
\begin{IEEEproof}
Appendix~\ref{app:lem:kappas}.
\end{IEEEproof}
We remark that the case 2) does not occur when $D \in \{ 2, 3\}$. 

From Lemma \ref{lem:kappas}, when the sample complexity $n$ satisfies~\eqref{eqn:sam_suff1} we can further bound $P_e^{(k)}$ in~\eqref{eqn:Sigma_s bound} by
\beq 
\begin{aligned}
P_e^{(k)}&\leq 2\sum_{s \leq \frac{k}{2}} \left(k^{-s}+2^{-2\sqrt{k}}+2^k e^{-k}\right)\\
&\leq c'\left(\frac{1}{k-1}+k2^{-2\sqrt{k}}+k2^ke^{-k}\right)
\end{aligned}
\eeq
for some constant $c'>0$. Note that this upper bound converges to 0 as $k\to\infty$. 

\subsection{Proof of~Theorem \ref{thm:main2}: Necessary Conditions for Almost Exact Recovery}\label{sec:proof2}
Theorem \ref{thm:main2} asserts that to reliably recover a fixed proportion $(1-\delta)k$ of variables for $\delta=o(1)$ from the parity-based querying strategy with average query difficulty $\bard$, it is necessary that 
$
n\geq (1-\epsilon)\cdot\max\left\{k(1-\delta-H_{\sf B}(\delta)),\frac{k\log (1/\delta) }{\bar{d}}\right\}
$
for any $\epsilon>0$ independent of $k,\delta$ and $\bard$. We show this result by proving the two lower bounds in~\eqref{eqn:lower_al} separately.

In Lemma~\ref{lem:sep1} below, by using Fano's inequality~\cite{cover2012elements} we show that to make $P_{e,\alpha}^{(k)}\to 0$ as $k\to\infty$ it is necessary that $n\geq (1-\epsilon)k(1-\delta-H_{\sf B}(\delta))$. A slightly tighter bound can also be proved by using the converse of the rate-distortion theorem~\cite{csiszar2011information}, but the difference between the two bounds is just  $O(\delta)$, which is negligible when $\delta=o(1)$.
\begin{lemma}\label{lem:sep1} 
{\it
To guarantee $P_{e,\alpha}^{(k)}\to 0$ as $k\to\infty$, it is necessary that
\beq
n\geq(1-\epsilon) k(1-H_{\sf B}(\delta)-\delta)
\eeq
for any $\epsilon>0$ independent of $k$ and $\delta$.}
\end{lemma}   
\begin{IEEEproof}
We use Fano's inequality to prove this. Define an indicator of the error event:
\beq
\mathbb{1}_E=
\begin{cases}
0,\quad\text{when } A(\bx,\hat{\bx}(\by))\geq \alpha ,\\
1,\quad\text{when }A(\bx,\hat{\bx}(\by))< \alpha .
\end{cases}
\eeq
The conditional entropy $H(\bx,\mathbb{1}_E|\hat{\bx})$ can be bounded above as
\beq
\begin{split}\label{eqn:firstFano}
H(\bx,\mathbb{1}_E|\hat{\bx})&= H(\mathbb{1}_E|\hat{\bx}) + H(\bx| \mathbb{1}_E,\hat{\bx})\\
&\leq \log 2+ (1- P_{e,\alpha}^{(k)})\cdot H(\bx| \mathbb{1}_E=0,\hat{\bx})
+P_{e,\alpha}^{(k)} \cdot H(\bx| \mathbb{1}_E=1,\hat{\bx}).
\end{split}
\eeq
for $P_{e,\alpha}^{(k)}$ in~\eqref{eqn:alerr_prob}.
When $\mathbb{1}_E=0$, $\bx$ and $\hat{\bx}$ have at least $\alpha k$ common symbols.  
Define a random variable $Z\in\left\{1,\dots,{ k \choose \alpha k}\right\}$ that specifies the location of those $\alpha k$ common symbols in the length-$k$ vectors $\bx$ and $\hat{\bx}$.
Note that
\beq
\begin{split}\label{eqn:bd_1E0}
H(\bx| \mathbb{1}_E=0,\hat{\bx})&\leq  H(\bx,Z| \mathbb{1}_E=0,\hat{\bx}) \\
&= H(Z| \mathbb{1}_E=0,\hat{\bx})+ H(\bx| \mathbb{1}_E=0,\hat{\bx}, Z)\\
&\leq \log {k \choose \alpha k} + (1-\alpha)k\\
&\leq kH_{\sf B}(\alpha)+(1-\alpha)k +o(k)
\end{split}
\eeq
where $H_{\sf B}(\alpha)=-\alpha\log\alpha-(1-\alpha)\log(1-\alpha)$.
By using~\eqref{eqn:bd_1E0}, we can further bound $H(\bx,\mathbb{1}_E|\hat{\bx})$ in~\eqref{eqn:firstFano} by
\beq
\begin{split}
&H(\bx,\mathbb{1}_E|\hat{\bx})\\
&\leq \log 2 +(1- P_{e,\alpha}^{(k)})\left( kH_{\sf B}(\alpha)+(1-\alpha)k\right)
+ P_{e,\alpha}^{(k)} \cdot k+o(k)\\
&= \log 2 + k\left(H_{\sf B}(\alpha)+(1-\alpha)+P_{e,\alpha}^{(k)} (\alpha - H_{\sf B}(\alpha))\right)
+o(k).
\end{split}
\eeq

By using this upper bound and the data processing inequality for $\bx\to \by\to \hat{\bx}$, we can show that
\beq
\begin{split}
k&=H(\bx)\\
&=H(\bx|\hat{\bx})+I(\bx;\hat{\bx})\\
&\leq H(\bx, \mathbb{1}_E|\hat{\bx})+I(\bx;\by)\\
&\leq \log 2 + k\left(H_{\sf B}(\alpha)+(1-\alpha)+P_{e,\alpha}^{(k)} (\alpha - H_{\sf B}(\alpha))\right)
+n+o(k).
\end{split}
\eeq
By rearranging terms, 
\beq
P_{e,\alpha}^{(k)}\geq \frac{k(\alpha - H_{\sf B}(\alpha))-n-\log 2-o(k)}{k(\alpha - H_{\sf B}(\alpha))}.
\eeq
For any constant $\epsilon>0$ independent of $k$, if 
\beq
n< (1-\epsilon)k(\alpha - H_{\sf B}(\alpha)),
\eeq
then $P_{e,\alpha}^{(k)}>c>0$ for some positive constant $c>0$.

Therefore, to make $P_{e,\alpha}^{(k)}\to 0$, it is necessary that 
\beq
n\geq (1-\epsilon)k(\alpha - H_{\sf B}(\alpha))
\eeq
for any constant $\epsilon>0$ independent of $k$.
\end{IEEEproof}

 In Lemma~\ref{lem:sep2} below, by using the fact that $P_{e,\alpha}^{(k)}$ can never tend to 0 if more than $\delta k$ input nodes are isolated, we show that the sample complexity $n$ should satisfy $n\geq (1-\epsilon)\frac{k\log (1/\delta) }{\bar{d}}$ for any $\epsilon>0$ independent of $k$.

\begin{lemma}\label{lem:sep2}
{\it
To guarantee $P_{e,\alpha}^{(k)}\to 0$ as $k\to\infty$ for the parity-based querying with query-degree distribution $(\Omega_1,\dots,\Omega_k)$ satisfying $\bard=\sum_{d=1}^k d\Omega_d$ and $\sum_d d^2 \Omega_d \ll k \bard$, it is necessary that
\beq\label{eqn:Lem5main}
n\geq (1-\epsilon)\frac{k\log (1/\delta) }{\bar{d}}
\eeq
for any $\epsilon>0$ independent of $k$, $\delta$ and $\bard$, where the average query difficulty $\bard=O(\log k)$.
}
\end{lemma}
\begin{IEEEproof}
The proof uses the second moment method to establish the necessary condition for almost exact recovery of $\alpha k$ variables with high probability where $\alpha=1-\delta$ for some $\delta=o(1)$ with $\delta\geq 1/k$.
In the random bipartite graph in Fig.~\ref{fig:graph}, if more than $(1-\alpha) k=\delta k$ input nodes are isolated, then it is impossible to reliably recover $\alpha k$ or more of the input nodes with high probability, i.e., the error probability is $\Pr(A(\bx,\hat{\bx}(\by))< \alpha )>c'>0$ for some constant $c'>0$. 
We analyze the probability of this event when the average degree of each output node equals $\bard$. 
 More precisely, we fix 
\beq
	n_c = \frac{ck\log (1/\delta) }{\bar{d}},
\eeq
for some constant $0<c<1$ independent of $k$, and show that if $n = n_c$ then 
\[
	\Pr(\text{more than } \delta k \text{ input nodes are isolated}) \to 1 \quad \text{ as } k \to \infty.
\] 
Since the number of isolated input nodes increases as the number of output nodes $n$ decreases for a fixed query difficulty $\bard$, it implies that it is necessary to have $n \geq n_c$ to guarantee the almost exact recovery of $\alpha$ fraction. We assume that $\bard = O(\log k)$, since when $\bard\gg \log k$ Lemma~\ref{lem:sep1} gives a tighter necessary condition on $n$ than~\eqref{eqn:Lem5main}. 

The assumed condition
\beq \label{eq:d^2_assumption}
	\sum_d d^2 \Omega_d \ll k \bard,
\eeq
means that $\Omega_d$ for $d = \Theta(k)$ makes negligible contribution to the mean $\bard$. Note that
\beq
	\sum_d d^2 \Omega_d \leq \sum_d kd \Omega_d = k \bard.
\eeq

For each input node $i$ ($1 \leq i \leq k$), we define a random variable $Z_i$ as
\beq
	Z_i = \begin{cases}
		0 & \text{ if the $i$-th input node is isolated}, \\
		1 & \text{ if the $i$-th input node is connected}.
	\end{cases}
\eeq
As computed in \eqref{eqn:iso_pro},
\beq
	\Pr(Z_i = 0) = \left( 1- \frac{\bard}{k} \right)^{n_c}, \qquad \Pr(Z_i = 1) = 1-\left( 1- \frac{\bard}{k} \right)^{n_c}.
\eeq
Let
\beq
	Z = Z_1 + \dots + Z_k.
\eeq
By definition, $Z$ is the number of connected input nodes. We also have
\beq \label{eq:EX_1}
	\E[Z] = k \E[Z_1] = k \left( 1-\left( 1- \frac{\bard}{k} \right)^{n_c} \right) = (1+o(1)) k (1-e^{-\frac{n_c \bard}{k}}) = (1+o(1)) k (1-\delta^c).
\eeq
For a more precise estimate in \eqref{eq:EX_1}, note that
\beq
	\frac{\bard}{k} \leq -\log \left(1-\frac{\bard}{k} \right) = \int_{1- \frac{\bard}{k}}^1 \frac{1}{t} \, dt \leq \frac{\bard}{k} \frac{1}{1-(\bard/k)} = \frac{\bard}{k-\bard},
\eeq
hence
\beq
	e^{-\frac{n_c \bard}{k-\bard}} \leq \left( 1-\frac{\bard}{k} \right)^{n_c} \leq e^{-\frac{n_c \bard}{k}}.
\eeq
From the first inequality above, we also find that
\beq
	e^{-\frac{n_c \bard}{k}} \leq \left( 1-\frac{\bard}{k} \right)^{\frac{n_c (k-\bard)}{k}} = \left( 1-\frac{\bard}{k} \right)^{n_c} \left( 1-\frac{\bard}{k} \right)^{-\frac{n_c \bard}{k}} = \left( 1+ O \left(\frac{n_c \bard^2}{k^2} \right) \right) \left( 1-\frac{\bard}{k} \right)^{n_c}.
\eeq
Thus, we obtain an estimate
\beq \begin{split} \label{eq:E_estimate}
	\left( 1-\frac{\bard}{k} \right)^{n_c} &= \left( 1+ O \left(\frac{n_c \bard^2}{k^2} \right) \right) e^{-\frac{n_c \bard}{k}} = \left( 1+ O \left(\frac{n_c \bard^2}{k^2} \right) \right) \delta^c \\
	&= \left( 1+ O \left(\frac{\bard \log(1/\delta)}{k} \right) \right) \delta^c = \left( 1+ O \left(\frac{(\log k)^2}{k} \right) \right) \delta^c,
\end{split} \eeq
where in the last equality we used the assumptions that $\bard = O(\log k)$ and $\delta \geq \frac{1}{k}$.

To estimate $\Pr(Z>\alpha k)$, we compute $\var(Z)$ and apply Markov's inequality. Since
\beq\label{eqn:helpZ}
	\E[Z^2] = \E\left[ \left( \sum_{i=1}^k Z_i \right)^2 \right] = k \E[Z_1^2] + k(k-1) \E[Z_1 Z_2],
\eeq
we need to find $\E[Z_1 Z_2]$. Since $Z_1$ and $Z_2$ are $0$ or $1$, $\E[Z_1^2] = \E[Z_1]$ and
\beq \begin{split}
	\E[Z_1 Z_2] &= \Pr(Z_1 Z_2 = 1) = \Pr(Z_1=1, Z_2=1) = \Pr(Z_1=1) - \Pr(Z_1=1, Z_2=0) \\
	&= \Pr(Z_1=1) - \big( \Pr(Z_2=0) - \Pr(Z_1=0, Z_2=0) \big)  \\
	&= 1- 2\left( 1- \frac{\bard}{k} \right)^{n_c} + \Pr(Z_1=0, Z_2=0).
\end{split} \eeq
For a fixed output node with degree $d$, the probability that the edges are not connected to the input nodes $1$ and $2$ is
\beq
	\binom{k-2}{d} \big/ \binom{k}{d} = \frac{(k-2)! (k-d)!}{k! (k-2-d)!} = \frac{(k-d)(k-d-1)}{k(k-1)}
\eeq
and since $d$ follows a distribution $(\Omega_1,\dots, \Omega_k)$,
\beq
	\sum_{d=1}^k \Omega_d \cdot \frac{(k-d)(k-d-1)}{k(k-1)} = \frac{k(k-1) - \bard(2k-1) + \sum_d \Omega_d \cdot d^2}{k(k-1)}.
\eeq
Since output nodes are generated independently, the probability that the input nodes 1 and 2 are not connected to any of $n_c$ output nodes is
\beq
\left(\frac{k(k-1) - \bard(2k-1) + \sum_d \Omega_d \cdot d^2}{k(k-1)}\right)^{n_c}.
\eeq
Thus, we obtain that
\beq \begin{split} \label{eq:EX_1X_2}
	\E[Z_1 Z_2] &= 1- 2\left( 1- \frac{\bard}{k} \right)^{n_c} + \left( 1- \frac{(2k-1)\bard - \sum_d \Omega_d \cdot d^2}{k(k-1)} \right)^{n_c} \\
	&= 1 - 2\delta^c + \delta^{2c} +o(\delta^{2c}) +  O \left(\frac{(\log k)^2 \delta^c}{k} \right),
\end{split} \eeq
where we apply the estimate \eqref{eq:E_estimate} and the assumption \eqref{eq:d^2_assumption}.

We now have from \eqref{eqn:helpZ}, \eqref{eq:EX_1} and \eqref{eq:EX_1X_2}
\beq \begin{split}
	\E[Z^2] &= k \left( 1-\left( 1- \frac{\bard}{k} \right)^{n_c} \right) \\
	&\quad+ k(k-1) \left( 1- 2\left( 1- \frac{\bard}{k} \right)^{n_c} + \left( 1- \frac{(2k-1)\bard - \sum_d \Omega_d \cdot d^2}{k(k-1)} \right)^{n_c} \right) \\
	&= k(1-\delta^c) + (k^2 -k) ( 1-2\delta^c + \delta^{2c}) + o(k^2 \delta^{2c}) + O\left( k (\log k)^2 \delta^c \right) \\
	&= k^2 (1-\delta^c)^2 + O(k \delta^c) + o(k^2 \delta^{2c}) + O\left( k (\log k)^2 \delta^c \right).
\end{split} \eeq
Note that $\delta \geq \frac{1}{k}$, hence $k \delta^c \ll k (\log k)^2 \delta^c \ll k^2 \delta^{2c}$ for any $0<c<1$. We also have from \eqref{eq:EX_1} and \eqref{eq:E_estimate} that
\beq
	(\E[Z])^2 = k^2 (1-\delta^c)^2 + O\left( k (\log k)^2 \delta^c \right)
\eeq
hence
\beq
	\var(Z) = \E[Z^2] - (\E[Z])^2 = o(k^2 \delta^{2c})
\eeq

Therefore, from Markov's inequality,
\beq
	\Pr(Z \geq \alpha k) = \Pr(Z - \E[Z] \geq \alpha k - \E[Z]) \leq \frac{\var(Z)}{(\alpha k - \E[Z])^2} = O\left( \frac{\var(Z)}{(k \delta^c)^2} \right) = o(1).
\eeq
This shows that $\Pr(\text{more than } \delta k \text{ input nodes are isolated}) \to 1$ as $k \to \infty$ if $n=n_c$.
\end{IEEEproof}

\subsection{Proof of~Theorem \ref{thm:main3}: Sufficient Conditions for Almost Exact Recovery}\label{sec:proof3}
Theorem \ref{thm:main3} asserts that by using the specified parity-based querying strategy, with average query degree $\bard>1$, at least $\alpha k=(1-\delta) k$  variables can be reliably recovered, i.e., $P_{e,\alpha}^{(k)}\to0$ as $k\to\infty$ for $\delta=o(1)$,  with sample complexity
$n\geq c\cdot \max\left\{k,\frac{k\log (1/\delta)}{\bard -1}\right\}$
for some constant $c>0$ where $\bard>1$. If $\bard=1$, it is sufficient to have the sample complexity $n=(1+\epsilon)k\log(1/\delta)$ for any $\epsilon>0$. 

The result for $\bard=1$ can be easily established by relating it to the classical coupon collecting problem, or equivalently the process of throwing balls randomly into bins where the balls are considered as edges in the graph in Fig.~\ref{fig:graph} and the bins are considered as input nodes. A sketch of the proof is as follows.
When $\Omega_1=1$ (and thus $\bard=1$), since every output node  is connected to a unique input node, if more than $(1-\delta)k$ input nodes are connected by any of the output nodes then we can guarantee the recovery of $(1-\delta)k$ input nodes. Therefore, we want to find the number of balls, which is equal to $n$ for $d=1$, to fill $(1-\delta)k$ bins when the balls are thrown uniformly at random into $k$ bins. 
When $(i-1)$ bins are already filled, the expected number of balls required to fill the $i$-th new bin is $k/(k-i+1)$. Since the expected number of balls to fill $(1-\delta)k$ different bins is $\sum_{i=1}^{(1-\delta)k} k/(k-i+1)\to k\log(1/\delta)$ as $k\to\infty$, if we have $n$ balls slightly larger than $k\log(1/\delta)$ then we can fill the $(1-\delta)k$ bins with high probability. 

We next consider the case $\bard>1$. 
If $\delta \leq k^{-\frac{1}{2}+\epsilon}$ for some $\epsilon \in (0, \frac{1}{2})$, then for 
$$
	n \geq \frac{c k \log (1/\delta)}{\bard} \geq c\left(\frac{1}{2} -\epsilon \right) \frac{k \log k}{\bard},
$$
recovering $\alpha k$ symbols is trivially guaranteed by the exact recovery conditions in Theorem \ref{thm:main}. Thus, we consider the case $\delta > k^{-\frac{1}{2}+\epsilon}$ only.

We first prove that it is sufficient to have $n \geq 2k$ if $\bard = \log(1/\delta) + O(1)$
 when the query degree is sampled from the soliton distribution~\eqref{eqn:modSol} with maximum degree $D=1/\delta$. For this distribution, the average degree is $\bard=\log(1/\delta)+O(1)$. 

We consider the BP decoding algorithm, explained in Section \ref{sec:IIIB}, as a decoding rule for the parity queries.
As a recap, we say that an input node is uncovered if its value is unknown. 
At the first step all output nodes of degree one are released and their unique neighboring input nodes are covered. 
At each subsequent step the decoder processes a covered input node and then removes it (with all its edges) from the graph. 
Processing an input node may {release} output nodes that subsequently have exactly one remaining neighboring input node. 
The newly released output node may increase the number of covered input node if its unique neighboring input node has not yet been covered.
The BP decoder repeats this process until there remain no more uncovered input node, or there remain no more covered input nodes that have not yet been processed. 
If the number of covered input nodes that have not yet been processed does not drop to 0 until $\alpha k$ input nodes are processed, the decoding process is successful for the almost exact recovery of $\alpha$ fraction.

We use the following notations from~\cite{luby2002lt}.
 Let $q(d, L)$ be the probability that an output node of degree $d$ is released when $L$ input nodes remain unprocessed in the BP decoding process. Then, it is easy to check the following:
\begin{itemize}
\item $q(1, k) = 1$
\item For $d=2, \dots, k$ and for all $L = k-d+1, \dots, 1$,
\beq
	q(d, L) = \frac{d(d-1) \cdot L \cdot \prod_{j=0}^{d-3} \big( k- (L+1) -j \big)}{\prod_{j=0}^{d-1} (k-j)}.
\eeq
\item For all other $d$ and $L$, $q(d, L) = 0$.
\end{itemize}
Let $r(d,L)$ be the probability that an output node is chosen to have degree $d$ and is released when $L$ input nodes remain unprocessed. Let $r(L)$ be the overall probability that an output is released when $L$ input nodes remain unprocessed. 
\beq\label{eqn:defrL}
	r(d, L) = \Omega_d \cdot q(d, L), \qquad r(L) = \sum_{d=1}^k r(d, L).
\eeq
We remark that we may stop the decoding when $\delta k$ input nodes remain unprocessed, which means that we only consider $L = k, k-1, \dots, \delta k$. In the following lemma, we find a lower bound on $r(L)$.
\begin{lemma}\label{lem:rL}
{\it
For $L=k, k-1, \dots, \delta k$, we have
\beq\label{eqn:rLlower}
	r(L) \geq \frac{C_e}{k}
\eeq
for a constant $\frac{1}{2} < C_e < 1-e^{-1}$ where $\delta > k^{-\frac{1}{2}+\epsilon}$.}
\end{lemma}
\begin{IEEEproof}
Appendix~\ref{app:lem:rL}
\end{IEEEproof}

When $L$ input nodes remain unprocessed, let $\rho(L)$ be the number of unprocessed but covered input nodes. 
For the BP decoding process to be successful for almost exact recovery of $\alpha=(1-\delta)$ fraction, $\rho(L)$ should not drop to 0 for $L=k,k-1,\dots,\delta k$. 
To guarantee this, we try to keep the size of $\rho(L)$ stable during $L=k,k-1,\dots, \delta k$. 

First we check the size of $\rho(L)$ at the beginning of the decoding process, $L=k$. We set $n=2k$ and $R=k^{(1+\epsilon)/2}$. 
The expected number of output nodes with degree one is $n \Omega_1 = 2 \delta k$. Hence, the actual number of output nodes with degree one is at least $\delta k$ with high probability. With these nodes, since $\delta k \geq k^{\frac{1}{2} +\epsilon} \gg k^{\frac{1}{2} + \frac{\epsilon}{2}} \log k \geq R \log R$, we can apply the result from the classical coupon collecting problem to show that the initial number of the covered input nodes is $\omega(R)$. By filtering, we make $\rho(k)$ exactly equal to $2R$.

In the following lemma, we show that the fluctuations of $\rho(L)$ are less than $R$ for $L=k,k-1,\dots, \delta k$, so that we can guarantee that $\rho(L)$ does not vanish at least until $L=\delta k$.
\begin{lemma} \label{lem:stable_ripple}
{\it
When  $L$ input nodes remain unprocessed, the number $\rho(L)$ of covered input nodes that have not been processed satisfy $R \leq \rho(L) \leq 3R$ for $L=k, k-1, \dots, \delta k$ with high probability where $\delta > k^{-\frac{1}{2}+\epsilon}$.}
\end{lemma}
\begin{IEEEproof}
We prove the lemma inductively on $L$. By construction, the lemma holds for $L=k$ since we made $\rho(k) = 2R$. Suppose that the lemma holds for $L=k, k-1, \dots, \tilde L$ (for some $\tilde L > \delta k$). We will show that $R \leq \rho(\tilde L - 1) \leq 3R$. Here, instead of estimating $\rho(\tilde L) - \rho(\tilde L -1)$, we find an upper bound on $|\rho(k) - \rho(\tilde L -1)|$ by using the induction hypothesis.

Fix an integer $L' \in [\tilde L - 1, k-1]$ and consider the processing of an input node that leaves $L'$ input nodes unprocessed. Before processing there were $\rho(L'+1)$ unprocessed but covered input nodes, and the processing makes the number of the unprocessed covered input nodes decreased by one. From the induction hypothesis, we find that the number of the unprocessed covered input nodes is at most $(3R-1)$ before counting the increment of covered input nodes due to newly released output nodes.

For an output node to increase the number of the covered input nodes by one, it must be released and its only remaining edge must be connected to an input node that was previously uncovered. 
The probability that an output node is released when $L'$ input nodes remain unprocessed is equal to $r(L')$ in~\eqref{eqn:rLlower}.
When an output node is released and no other output nodes are released at the same time, the probability that this output node increases the number of the covered input nodes by one is
\beq
	\frac{L'- (\rho(L'+1)-1)}{L'},
\eeq
which is at least $(L'-3R)/L'$. However, if two or more output nodes are released simultaneously, then this probability may be lowered. To handle such a situation, we notice that the number of released output nodes is at most $R$ with high probability. (This can be checked by applying Chernoff bound together with the fact that the expected number of the released output nodes is $\Theta(1)$ since $\frac{C_e}{k} \leq r(L') < \frac{1}{k}$.) Hence, even when all output nodes that have been released by the processing of the same input node increase the number of covered input nodes by one, still the probability that a newly released output node further increases the number of the covered input nodes by one is at least $(L'-4R)/L'$. Note that $L' \geq \delta k \gg R$, hence $(L'-4R)/L' = 1+o(1)$. 

Let $p_{L'}$ be the probability with which an output node is released and makes the number of the covered input nodes increased by one when $L'$ input nodes remain unprocessed. We have found that
\beq\label{eqn:p_L'}
	p_{L'} \geq r(L') \cdot \frac{L'-4R}{L'} >\frac{C_e}{k} \cdot \frac{L'-4R}{L'} >\frac{1}{n},
\eeq
where the second inequality holds from Lemma \ref{lem:rL} and the last inequality is true for $n=2k$. By filtering, we make $p_{L'} = 1/n$. 

For each output node $i\in[1:n]$, we now define a random variable $Z_i$ such that $Z_i = 1$ if the $i$-th output node makes the number of the covered input nodes increased by one during which the number of unprocessed input nodes  decreases from $(k-1)$ to $(\tilde L -1)$, and $Z_i = 0$ otherwise. Then,
\beq\label{eqn:rhoL-1}
	\rho(\tilde L -1) = \rho(k) + \sum_{i=1}^n Z_i - (k- \tilde L + 1)=2R+ \sum_{i=1}^n Z_i - (k- \tilde L + 1).
\eeq
By construction, $Z_i$ is a Bernoulli random variable with 
\beq
	\Pr(Z_i = 1) = \sum_{L' = \tilde L -1}^{k-1} \frac{1}{n} = \frac{k-\tilde L}{n}.
\eeq
Using Chernoff bound, it can be proved that
\beq\label{eqn:Chernf_fluc}
	\Pr \left( \left| \sum_{i=1}^n Z_i - (k-\tilde L +1) \right| \geq R \right) \leq e^{-c k^{\epsilon}}
\eeq
for some constant $c>0$ independent of $k$. From~\eqref{eqn:rhoL-1} and~\eqref{eqn:Chernf_fluc}, we find that $R \leq \rho(\tilde L-1) \leq 3R$ with high probability.
\end{IEEEproof}

From Lemma \ref{lem:stable_ripple}, we immediately see that the BP decoding proceeds successfully  with high probability until $\alpha k=(1-\delta)k$ input nodes are processed, which implies that the almost exact recovery for $\alpha$ fraction is possible when $n=2k$ and $\bard=\log(1/\delta)+O(1)$.

In the case $\bard < \log(1/\delta)$, we show that the almost exact recovery is possible with $n = \frac{2k\log(1/\delta)}{\bard-1}(1+o(1))$. For the same $D=1/\delta$ and a given $\bard$, we adjust the soliton distribution~\eqref{eqn:modSol} to
\beq\label{eqn:adj_Sol}
\begin{aligned}
\Omega_d' = 
	\begin{cases}
	1 - \eta +  \frac{\eta}{D} & \text{ if } d=1 \\
	\frac{\eta}{d(d-1)} & \text{ if } 2\leq d \leq D \\
	0 & \text{ if } d > D,
	\end{cases}
\end{aligned}
\eeq
for $\eta$ which satisfies 
\beq
\bard=\sum_d d \Omega_d' = 1-\eta+{\eta}\sum_{i=1}^D\frac{1}{d}. 
\eeq
Define
\beq\label{eqn:Euler}
\gamma(D):=-\log D+\sum_{d=1}^{D}\frac{1}{d}=\int_{1}^{D}\left(-\frac{1}{x}+\frac{1}{\lfloor{x}\rfloor}\right)dx.
\eeq
As $D$ increases, $\gamma(D)$ increases and it converges to Euler--Mascheroni constant which is smaller than 0.58.
Therefore, we have
\beq
\eta=\frac{\bard-1}{\log D-1+\gamma(D)}=\frac{\bard-1}{\log(1/\delta)}(1+o(1)).
\eeq

For the adjusted soliton distribution, following the same idea as in Lemma \ref{lem:rL}, we can show that the probability $r(L)$ that an output is released when $L$ input nodes remain unprocessed satisfies
\beq
	r(L) \geq \eta \cdot \frac{C_e}{k}.
\eeq
Then, when $L$ input nodes remain unprocessed the probability $p_{L}$  that an output node is released and it increases the number of covered input nodes by one satisfies
\beq\label{eqn:pLnew}
p_L\geq r(L) \cdot \frac{L-4R}{L} \geq \eta \cdot \frac{C_e}{k}\cdot\frac{L-4R}{L}
\eeq
as in~\eqref{eqn:p_L'}. When we choose the number of output nodes to be $n=2k /\eta$, $p_L$ can be further bounded by
\beq
p_L \geq \eta \cdot \frac{C_e}{k}\cdot\frac{L-4R}{L}\geq \frac{1}{n}.
\eeq
Thus, by filtering, we can make $p_L = 1/n$. We can then use Chernoff bound to show that $\rho(L)$ is between $R$ and $3R$ with high probability for $L=k,k-1,\dots, \delta k$ by using similar arguments as in Lemma \ref{lem:stable_ripple}. This guarantees the almost exact recovery we desired for any $\bard<\log(1/\delta)$ with $n=2k/\eta=\frac{2k\log(1/\delta)}{\bard-1}(1+o(1))$.

\section{Numerical Experiments}
\label{sec:sim}
\subsection{Exact Recovery}
\begin{figure}[t]
\centerline{\includegraphics[width=\columnwidth]{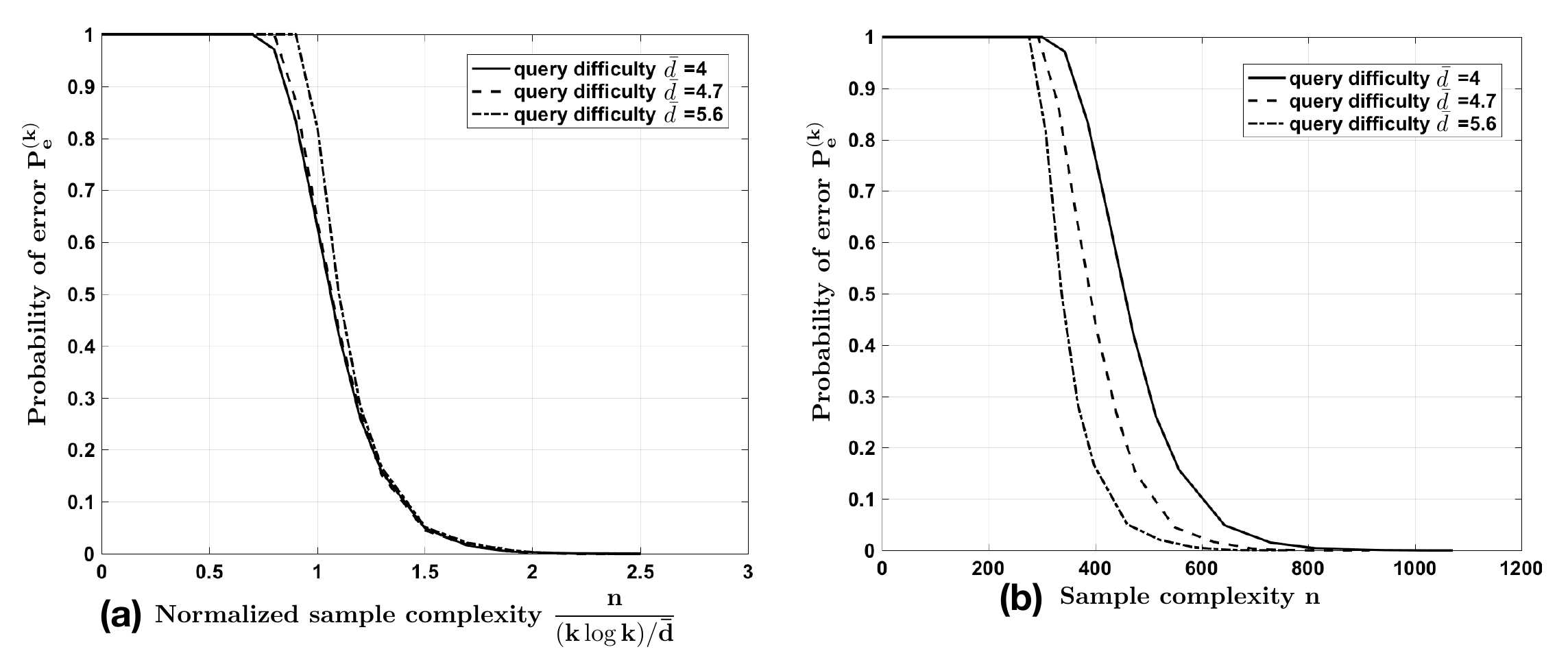}}
\caption{(a) Monte Carlo simulation (5000 runs) of the probability of error $P_e^{(k)}$ for exact recovery with $k=300$ for three different $\bard$'s (the query difficulties). The sample complexity is normalized by $(k\log k)/\bard$. 
We can observe the phase transition for $P_e^{(k)}$ around the normalized sample complexity equal to 1 for all the three query difficulties considered. (b) Same simulation results except that the horizontal axis is the un-normalized sample complexity. As the query difficulty increases, the sample complexity to make $P_e^{(k)}$ close to 0 decreases. This illustrates the trade-offs between the query difficulty and the sample complexity. }
\label{fig:PeMC}
\end{figure}

In this section, we illustrate the tightness of~Theorem \ref{thm:main} by providing empirical performance analysis for the probability of error in the recovery of $k$ binary variables as a function of the sample complexity and query difficulty, where parity-based queries are generated with soliton degree distribution \eqref{eqn:modSol} with query difficulty $\bard$.
Fig.~\ref{fig:PeMC}-(a) shows Monte Carlo simulation results for the probability of error $P_e^{(k)}$, defined in~\eqref{eqn:err_prob}, where the number of binary variables to be recovered is fixed at $k=300$.
We plot $P_e^{(k)}$ in terms of the normalized sample complexity, normalized by $(k\log k)/\bard$ where $\bard$ is the query difficulty. We run the simulations for three different query difficulties, $\bard=$4, 4.7, 5.6, which correspond to $D=30,60,130$, respectively, where $D$ is the maximum degree of the output node (the maximum size of the subset of variables per query).

Observe the phase transition of $P_e^{(k)}$ as a function of sample complexity $n$ that occurs in the vicinity of $n=1$. Theorem \ref{thm:main} states that, with sample complexity of $c_u\cdot \max\{k,(k\log k)/\bard\}$, for some constant $c_u>0$, $P_e^{(k)}\to 0$ as $k\to\infty$ is guaranteed. The simulation results show that $c_u\approx 1$ is sufficient to produce a dramatic decrease of $P_e^{(k)}$. 
The figure demonstrates the trade-offs between the query difficulty and the sample complexity. Specifically, the number of measurements to the parity-based queries that is required to reliably recover $k$ binary variables is inversely proportional to the query difficulty when $\bard = O(\log k)$. 
Note that for the soliton distribution~\eqref{eqn:modSol}, the query difficulty is $O(\log k)$, and thus $\max\{k,(k\log k)/\bard\}=\Theta( (k\log k)/\bard)$. 
In Fig.~\ref{fig:PeMC}-(b), we show the same simulation with un-normalized sample complexity indexing the horizontal axis. From this plot, we can observe that as the query difficulty increases, the required number of measurements to make $P_e^{(k)}$ close to 0 decreases.


\subsection{Almost Exact Recovery}

\begin{figure}[t]
\centerline{\includegraphics[width=\columnwidth]{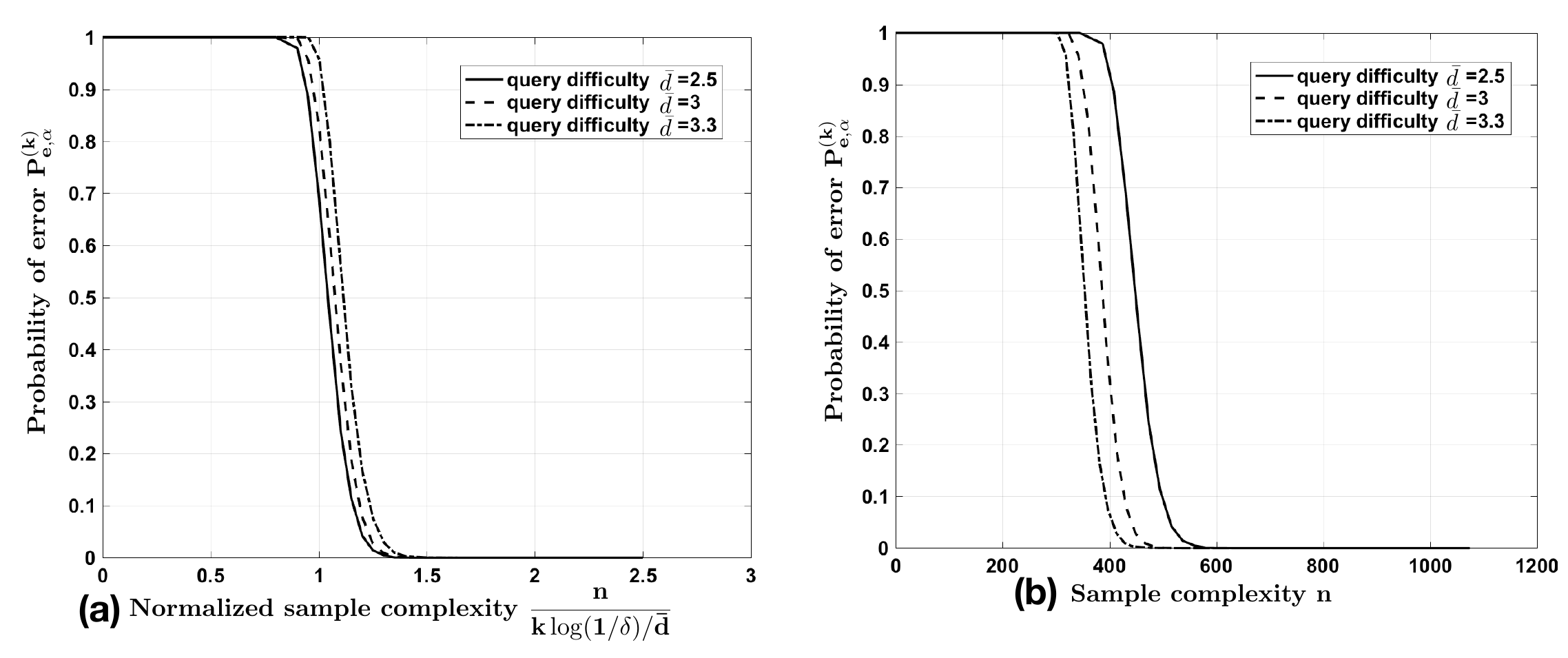}}
\caption{(a) Monte Carlo simulation (5000 runs) of the probability of error $P_{e,\alpha}^{(k)}$ for almost exact recovery with $\alpha=0.97$ and $k=300$ for three different $\bard$'s (the query difficulties). The sample complexity is normalized by $(k\log (1/\delta))/\bard$. 
We can observe the phase transition for $P_{e,\alpha}^{(k)}$ around the normalized sample complexity equal to 1 for all the three query difficulties considered.
(b) Same simulation results except that the horizontal axis is the un-normalized sample complexity. As the query difficulty increases, the sample complexity to make $P_{e,\alpha}^{(k)}$ close to 0 decreases. This illustrates the trade-offs between the query difficulty and the sample complexity for almost exact recovery. }
\label{fig:PePrMC_fixed_alpha_norm}
\end{figure}

We next consider almost exact recovery illustrating the theory with empirical performance analysis for the probability error associated with recovery of a fraction $\alpha$ of variables. 
As shown in Theorem \ref{thm:main3}, the sufficient number of measurements to guarantee reliable recovery of $\alpha k$ or more of the variables is
proportional to $\max\{k,\frac{k\log (1/\delta)}{\bard}\}$ for $\alpha=1-\delta$ with $\delta=o(1)$.  
This is supported by Monte Carlo simulation results for the probability of error $P_{e,\alpha}^{(k)}$, defined in~\eqref{eqn:alerr_prob}, for a fixed $k=300$, for several values of the recovery fraction parameter $\alpha$ and query difficulty $\bard$. Here, the output symbols are generated according to the soliton distribution~\eqref{eqn:modSol} and decoded by the BP decoding rule.


In Fig.~\ref{fig:PePrMC_fixed_alpha_norm}-(a), for a fixed $\alpha=0.97$ ($\delta=0.03$) we compare $P_{e,\alpha }^{(k)}$ for $\bard=2.5, 3, 3.3$, which correspond to $D=6,10,15$, respectively, where $D$ is the maximum degree of the output node. For these values of $\bard$ and $\delta$, we can check that $\log(1/\delta)>\bard\approx \log D$, and thus the required number of measurements to make $P_{e,\alpha }^{(k)}\to 0$ is $n\propto \frac{k\log (1/\delta)}{\bard}$, as stated in Theorem \ref{thm:main3}. In Fig.~\ref{fig:PePrMC_fixed_alpha_norm}-(a), we plot $P_{e,\alpha }^{(k)}$'s for the normalized sample complexity, $n/(\frac{k\log(1/\delta)}{\bard})$, and show that $P_{e,\alpha }^{(k)}$ drops to 0 in the vicinity of normalized sample complexity equal to 1.
This result shows that the sample complexity is not only proportional to $\frac{k\log (1/\delta)}{\bard}$ but the actual constant factor is roughly equal to 1. 
In Fig.~\ref{fig:PePrMC_fixed_alpha_norm}-(b), we plot the same $P_{e,\alpha }^{(k)}$'s with un-normalized sample complexity indexing the horizontal axis.  We can observe that as the query difficulty increases, the minimum $n$ to make $P_e^{(k)}$ close to 0 decreases.

\begin{figure}[t]
\centerline{\includegraphics[scale=0.6]{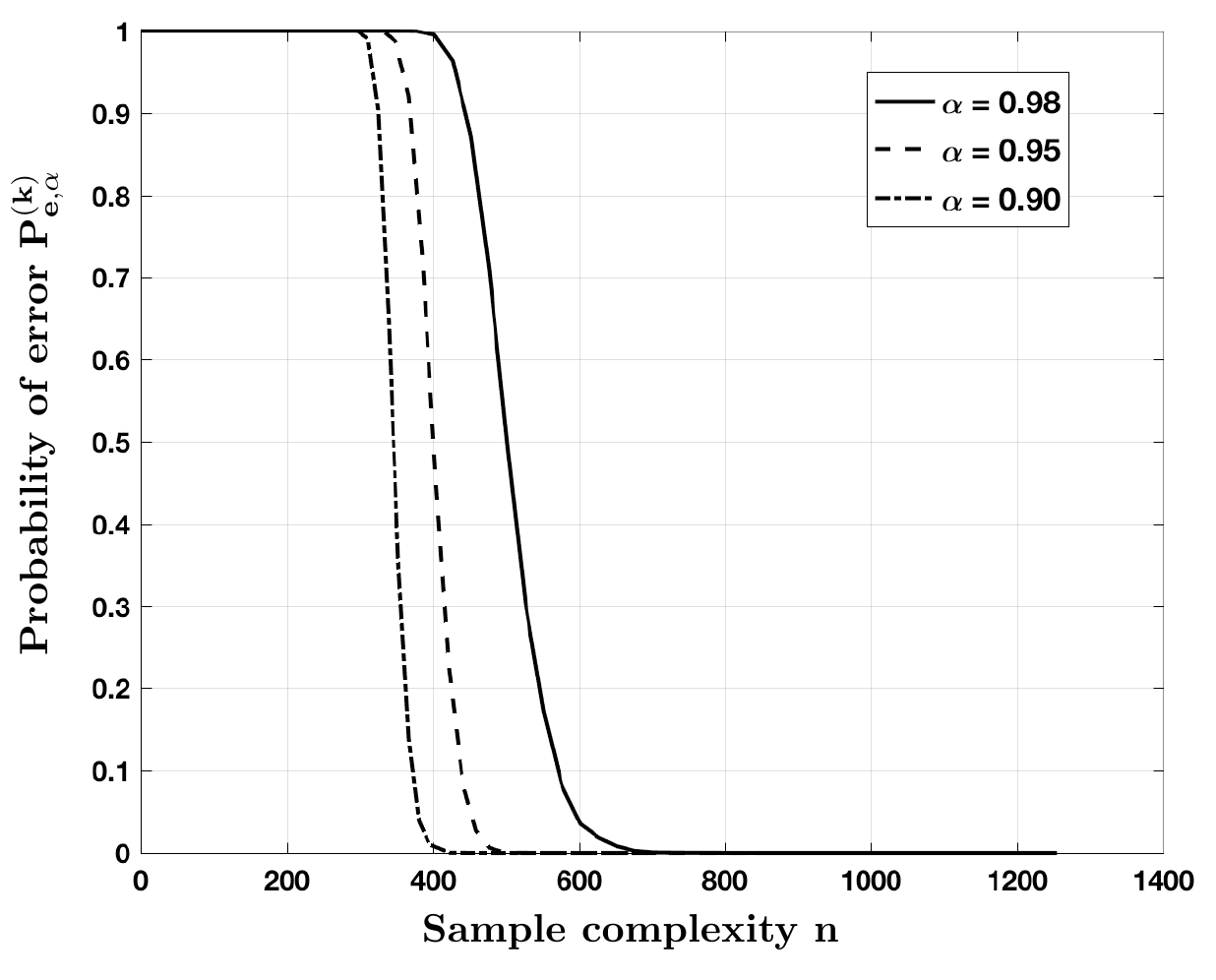}}
\caption{Monte Carlo simulation (5000 runs) of the probability of error $P_{e,\alpha}^{(k)}$ over sample complexity $n$ for almost exact recovery with a fixed $\bard=2.5$ and $k=300$ for three different $\alpha$'s (the fraction of recovery).  
The required $n$ to make $P_{e,\alpha}^{(k)}$ close 0 increases for larger $\alpha$. }
\label{fig:PePrMC_fixed_D_notnorm}
\end{figure}

In Fig.~\ref{fig:PePrMC_fixed_D_notnorm}, we observe the effect of $\alpha=1-\delta$ on the sample complexity $n$ for a fixed query difficulty $\bard=2.5$ (D=6). 
For the almost exact recovery,  we claim that the recovery is reliable when $\alpha k$ or more of the variables are recovered with high probability. 
We plot $P_{e,\alpha}^{(k)}$ for $\alpha=(0.9,0.95,0.98)$ for $k=300$ and show that the required $n$ to make $P_{e,\alpha}^{(k)}$ close 0 increases as the value of $\alpha$ increases. 

\section{Conclusions}
\label{sec:con}

We have proposed a method for designing a sequence of parity queries to recover the $k$ binary variables with high probability at the optimal sample complexity (up to constant factors) in a response model where only a randomly selected subset of the queries is answered. In particular, defining the query difficulty $\bard$ as the average size of the query subsets we analyzed the fundamental trade-offs between recovery accuracy $\alpha$, query difficulty $\bard$ and sample complexity $n$.
We considered two recovery conditions, exact recovery and almost exact recovery with $\alpha=1-\delta$ fraction for $\delta=o(1)$, and showed that for exact recovery it is necessary and sufficient to have the sample complexity $n= c_0\cdot \max\{k, (k\log k)/\bard\}$ and for almost exact recovery the sample complexity $n= c_1\cdot \max\{k, (k\log (1/\delta))/\bard\}$ for constants $c_0,c_1>0$. We also provided a query-degree distribution for which the BP decoding process can guarantee the almost exact recovery of $k$ variables at the optimal sample complexity.

There are several interesting future research directions related to this work. 
 One direction is to relax the assumption $\alpha \rightarrow 1$ so as to tolerate poor reconstruction of a larger fraction of variables. In particular, we can consider extending Theorem \ref{thm:main2} and Theorem \ref{thm:main3} on almost exact recovery to the partial recovery problem where some smaller fraction $\alpha k$ of the variables are to be recovered, e.g.,  $\alpha\in(1/2,1)$.
An interesting question is whether it is possible to recover a smaller number $\alpha k$ of variables with only $n=\Theta(k)$ measurements even with a very low query difficulty $\bard=\Theta(1)$, which does not increase in $k$. 
Recall that in the exact recovery problem the query difficulty $\bard=\Theta(1)$ required a sample complexity of $n=\Theta(k\log k)$.

Another interesting direction is to apply the proposed parity-based query design to a more general measurement model with applications in crowdsourcing systems where workers may provide incorrect answers. 
For real crowdsourced labeling problems the probability that a worker provides an incorrect answer changes depending on the query difficulty. Designing querying schemes that minimize the sample complexity in recovering the attributes of objects under such a noisy model is a worthwhile problem for future study.

\appendices
\section{Proof of Lemma \ref{lem:comb}}\label{app:lem:comb}
To prove this lemma, we refer to the similar bound provided in~\cite{ahna2019community}.
\begin{lemma}\label{lem:ahn}
{\it
Let $\beta=\ceil*{\max\left\{\frac{k-d+1}{2d+1},\frac{d+1}{2(k-d)+1}\right\}}$ and $\alpha=\max\left\{\frac{k-d+1}{d},\frac{d+1}{k-d}\right\}$. Then we have
\beq
\sum_{\substack{i\leq d\\ i \text{ is odd}}}{s\choose i}{k-s\choose d-i}\geq 
\begin{cases}
\frac{2s}{5\alpha}{k\choose d},& \text{when } s<\beta,\\
\frac{1}{5}{k\choose d}, & \text{when } \beta\leq s\leq k-\beta,\\
\frac{2(k-s)}{5\alpha}{k\choose d}, &\text{when }k-\beta<s.
\end{cases}
\eeq
}
\end{lemma}

Note that
\beq
{k\choose d}=\sum_{\substack{i\leq d\\ i \text{ is odd}}}{s\choose i}{k-s\choose d-i}+\sum_{\substack{i\leq d\\ i \text{ is even}}}{s\choose i}{k-s\choose d-i}.
\eeq
Therefore, by using Lemma \ref{lem:ahn}, we can find an upper bound on $\sum_{\substack{i\leq d\\ i \text{ is even}}}{s\choose i}{k-s\choose d-i}$ as a scaling of ${k \choose d}$ of the form
 \beq
 \begin{aligned}
&I_d=\sum_{\substack{i\leq d\\ i \text{ is even}}}{s\choose i}{k-s\choose d-i}\\
&\leq
 \begin{cases} \label{eqn:I_d original bound}
\left(1- \frac{2s}{5\alpha}\right){k\choose d},& \text{when } s<\beta,\\
\frac{4}{5}{k\choose d}, & \text{when } \beta\leq s\leq k-\beta,\\
\left(1-\frac{2(k-s)}{5\alpha}\right){k\choose d}, &\text{when }k-\beta<s.
 \end{cases}
 \end{aligned}
 \eeq
Define
\beq
\kappa(s) = \frac{k-s+1}{2s+1}.
\eeq
We first consider the case $s \leq \frac{k}{2}$ (i.e., $s \leq k-s$). Since $\beta$ attains its maximum $\ceil*{\frac{k}{3}}$ at $d=1$ or $d=k-1$, we find that
$$
\beta < \frac{k}{2} \leq k-s.
$$
Hence, $k-\beta > s$ and the last case in \eqref{eqn:I_d original bound} is irrelevant.

\begin{enumerate}
	\item For $d \leq \frac{k}{2}$ (or, $k-d \geq d$),
	$$
	\beta = \ceil*{\frac{k-d+1}{2d+1}}, \qquad \alpha = \frac{k-d+1}{d}.
	$$
	Note that $\frac{k-\kappa(s)+1}{2\kappa(s)+1} = s$. Since $\frac{k-d+1}{2d+1}$ is an decreasing function of $d$, if $d < \kappa(s)$ then $\beta > s$. Thus,
	\beq
	I_d \leq \begin{cases}
	\left(1- \frac{2s}{5\alpha}\right){k\choose d},& \text{when } d < \kappa(s), \\
	\frac{4}{5}{k\choose d}, & \text{when } d \geq \kappa(s).
	\end{cases}
	\eeq
		
	\item For $d > \frac{k}{2}$ (or, $k-d < d$),
	$$
	\beta = \ceil*{\frac{d+1}{2(k-d)+1}}, \qquad \alpha = \frac{d+1}{k-d}.
	$$
	Proceeding as above, we get
	\beq
	I_d \leq \begin{cases}
	\left(1- \frac{2s}{5\alpha}\right){k\choose d},& \text{when } d > k-\kappa(s), \\
	\frac{4}{5}{k\choose d}, & \text{when } d \leq k-\kappa(s).
	\end{cases}
	\eeq

\end{enumerate}
In the case $s > \frac{k}{2}$, we can obtain the bounds for $I_d$ simply by changing $s$ to $k-s$. 

\section{Proof of Lemma \ref{lem:sam_case1}}\label{app:lem:sam_case1}
When the maximum query degree $D\leq c_d\log k$ for some $0<c_d\leq 1$, the bound in~\eqref{eqn:Sigma_s bound} can be simplified as
\beq
\begin{split}
P_e^{(k)}&  \leq 2\sum_{s \leq \frac{k}{2}} { k \choose s} \exp\left(-n\left(\frac{2s}{5}  \sum_{d=1}^{\ceil*{\kappa(s)}-1} \frac{d\,\Omega_d}{k-d+1}    +\frac{1}{5}\sum_{d=\ceil*{\kappa(s)}}^D \Omega_d \right)\right)\\
&  \leq 2\sum_{s \leq \frac{k}{2}} { k \choose s} \exp\left(-n\left(\frac{2s}{5k}  \sum_{d=1}^{\ceil*{\kappa(s)}-1} {d\,\Omega_d}   +\frac{1}{5}\sum_{d=\ceil*{\kappa(s)}}^D \Omega_d \right)\right).
\end{split}
\eeq

When we plug in the sample complexity $n=c\frac{k\log k}{\sum_{d=1}^D d\Omega_d}$ or some $c>0$, the bound becomes
\beq
\begin{split}
P_e^{(k)} & \leq 2\sum_{s \leq \frac{k}{2}} { k \choose s}\exp\left(-\frac{2c}{5}s\log k\frac{ \sum_{d=1}^{\ceil*{\kappa(s)}-1} {d\,\Omega_d} }{\sum_{d=1}^D d\Omega_d}-\frac{c}{5} k\log k \frac{ \sum_{d=\ceil*{\kappa(s)}}^{D} {\,\Omega_d} }{\sum_{d=1}^D d\Omega_d} \right)\\
\end{split}
\eeq
When $1\leq s\leq \lfloor \frac{k}{2c_d\log k}\rfloor$, we can bound the term in the summation as
\beq
\begin{split}
 & { k \choose s}\exp\left(-\frac{2c}{5}\log k\left(s\frac{ \sum_{d=1}^{\ceil*{\kappa(s)}-1} {d\,\Omega_d} }{\sum_{d=1}^D d\Omega_d}+\frac{k}{2} \frac{ \sum_{d=\ceil*{\kappa(s)}}^{D} {\,\Omega_d} }{\sum_{d=1}^D d\Omega_d}\right) \right)\\
 &\leq  \exp(s\log k) \exp\left(-\frac{2c}{5}s\log k\right)\leq\left(\frac{1}{k^\epsilon}\right)^s
 \end{split}
\eeq
where the first inequality follows from ${k\choose s}\leq \exp(s\log k)$ and from $sd\leq \frac{k}{2}$ for $d\leq c_d\log k$ and $s\leq\lfloor \frac{k}{2c_d\log k}\rfloor$, and the second inequality is true for $c\geq (1+\epsilon)\frac{5}{2}$ for any constant $\epsilon>0$.

When $ \lfloor \frac{k}{2c_d\log k}\rfloor+1\leq s\leq k/2$, we can bound the term in the summation as
\beq
\begin{split}
&{ k \choose s}\exp\left(-\frac{2c}{5}s\log k\frac{ \sum_{d=1}^{\ceil*{\kappa(s)}-1} {d\,\Omega_d} }{\sum_{d=1}^D d\Omega_d}-\frac{c}{5} k\log k \frac{ \sum_{d=\ceil*{\kappa(s)}}^{D} {\,\Omega_d} }{\sum_{d=1}^D d\Omega_d} \right)\\
&\leq{ k \choose s}^{\frac{ \sum_{d=1}^{\ceil*{\kappa(s)}-1} {d\,\Omega_d} }{\sum_{d=1}^D d\Omega_d}}\exp\left(-\frac{2c}{5}s\log k\frac{ \sum_{d=1}^{\ceil*{\kappa(s)}-1} {d\,\Omega_d} }{\sum_{d=1}^D d\Omega_d}\right)\times\\
&\quad  { k \choose s}^{\frac{ \sum_{d=\ceil*{\kappa(s)}}^D {d\,\Omega_d} }{\sum_{d=1}^D d\Omega_d}} \exp\left(-\frac{c}{5c_d} k \frac{ \sum_{d=\ceil*{\kappa(s)}}^{D} {\,d \Omega_d} }{\sum_{d=1}^D d\Omega_d} \right)\\
&\leq  \exp\left(-\epsilon (\log 2)2 s\log k\frac{ \sum_{d=1}^{\ceil*{\kappa(s)}-1} {d\,\Omega_d} }{\sum_{d=1}^D d\Omega_d}\right)  \exp\left(-\epsilon (\log 2) k \frac{ \sum_{d=\ceil*{\kappa(s)}}^{D} {d\,\Omega_d} }{\sum_{d=1}^D d\Omega_d} \right)\\
&\leq \exp\left(-\epsilon(\log 2) k\right).
\end{split}
\eeq
where the first inequality is from $d\leq D\leq c_d\log k$, and the second inequality follows from ${k\choose s}\leq \exp(s\log k)$, ${k\choose s}\leq \exp(k\log 2)$, and  for $c\geq (5\log 2)(1+\epsilon)\geq \frac{5}{2}(1+\epsilon) $. The last inequality is from $2s\log k\geq k$ for  $\lfloor \frac{k}{2c_d\log k}\rfloor+1\leq s\leq k/2$ with $0<c_d\leq 1$.

\section{Proof of Lemma \ref{lem:kappas}}\label{app:lem:kappas}
In this lemma, we prove an upper bound on ${k\choose s}e^{-n\Sigma_s}$ where
\beq
\begin{split}
\Sigma_s &= \frac{1}{5} \sum_{d=\ceil*{\kappa(s)}}^{k-\ceil*{\kappa(s)}} \Omega_d + \\
&\frac{2s}{5} \left( \sum_{d=1}^{\ceil*{\kappa(s)}-1} \frac{d\,\Omega_d}{k-d+1} + \sum_{d=k-\ceil*{\kappa(s)}+1}^{k} \frac{(k-d)\Omega_d}{d+1} \right).
\end{split}
\eeq
For the soliton distribution
\beq
\begin{aligned}\nonumber
\Omega_d = 
	\begin{cases}
	\frac{1}{D} & \text{ if } d=1 \\
	\frac{1}{d(d-1)} & \text{ if } 2\leq d \leq D \\
	0 & \text{ if } d > D,
	\end{cases}
\end{aligned}
\eeq
we have the query difficulty
$$
\log (D+1) < \bard = \frac{1}{D} + \sum_{d=2}^{D} \frac{1}{d-1} = \sum_{d=1}^D \frac{1}{d} < \log D + 1.
$$
For simplicity, here we assume that $D \geq 2$.
Recall that
\beq
\kappa(s) = \frac{k-s+1}{2s+1},
\eeq
which is a decreasing function of $s$, and $\kappa(s) > 0$ for $s \leq \frac{k}{2}$.

\begin{enumerate}

\item If $\ceil*{\kappa(s)} > D$,
$$
\Sigma_s \geq \frac{2s}{5} \sum_{d=1}^{\ceil*{\kappa(s)}-1} \frac{d\,\Omega_d}{k-d+1} > \frac{2s}{5k} \sum_{d=1}^{D} d\,\Omega_d = \frac{2s \bar d}{5k}.
$$
Thus, if $n \bar d \geq 5 k \log k$,
$$
\binom{k}{s} e^{-n\Sigma_s} < k^s \exp \left( -\frac{2ns \bar d}{5k} \right) \leq k^s k^{-2s} = k^{-s}.
$$

\item If $4 \leq \ceil*{\kappa(s)} \leq D$, we first notice that
\beq \label{eq:s_kappa}
s < \frac{k-2}{7} \Leftrightarrow \kappa(s) > 3 \Leftrightarrow \ceil*{\kappa(s)} \geq 4.
\eeq
Thus $s \leq \frac{k-2}{7}$ and $\kappa(s) -1 = \frac{k-3s}{2s+1} \geq \frac{4k}{7(2s+1)} \geq \frac{4k}{21s}$. In this case,
\beq
\begin{aligned}
\Sigma_s &\geq \frac{2s}{5} \sum_{d=1}^{\ceil*{\kappa(s)}-1} \frac{d\,\Omega_d}{k-d+1} \\
&> \frac{2s}{5k} \sum_{d=2}^{\ceil*{\kappa(s)}-1} \frac{1}{d-1}\\
& > \frac{2s}{5k} \log (\ceil*{\kappa(s)}-1)\\
& \geq \frac{2s}{5k} \log \left( \frac{4k}{21s} \right).
\end{aligned}
\eeq
Moreover, since $\frac{k}{s} \geq 7$, if $n \geq Ck$ for some sufficiently large $C$, ($C\geq 68$ suffices)
\beq
\begin{aligned}
&n\Sigma_s \\
&\geq \frac{2Cs}{5} \log \left( \frac{4k}{21s} \right)\\
& \geq 4s \log \left( \frac{k}{s} \right) + \left( \frac{2C}{5} -4 \right) s \log 7 + \frac{2Cs}{5} \log \left( \frac{4}{21} \right)\\
& \geq 4s \log \left( \frac{k}{s} \right).
\end{aligned}
\eeq
From Stirling's formula, we also have that
$$
\sqrt{2\pi} n^{n+\frac{1}{2}} e^{-n} \leq n! \leq e n^{n+\frac{1}{2}} e^{-n},
$$
hence
\begin{equation} \begin{split}
\binom{k}{s} 
&\leq \frac{e k^{k+\frac{1}{2}} e^{-k}}{2\pi (k-s)^{k-s+\frac{1}{2}} e^{-(k-s)} s^{s+\frac{1}{2}} e^{-s}} \\
&\leq \frac{\sqrt{k}}{2\sqrt{(k-s)s}} \cdot \frac{k^k}{(k-s)^{k-s} s^s} \\
&\leq  \left( \frac{k}{s} \right)^s \left(1-\frac{s}{k} \right)^{s-k} \\
&\leq \left( \frac{k}{s} \right)^s e^{s-\frac{s^2}{k}} = \exp \left( s \log \left( \frac{k}{s} \right) + s - \frac{s^2}{k} \right) \\
&\leq \exp \left( 2s \log \left( \frac{k}{s} \right) \right).
\end{split} \end{equation}
Thus, if $n \geq 68k$,
\beq
\binom{k}{s} e^{-n\Sigma_s} \leq \exp \left( -2s \log \left( \frac{k}{s} \right) \right) = \left( \frac{k}{s} \right)^{-2s}.
\eeq
Note that
\beq
\left( \frac{k}{s} \right)^{-2s} \leq
	\begin{cases}
	k^{-s} & \text{ if } s \leq \sqrt k \,, \\
	2^{-2\sqrt k} &\text{ if } \sqrt k < s \leq k/2 \,.
	\end{cases}
\eeq
\item If $\ceil*{\kappa(s)} = 3$, we find from \eqref{eq:s_kappa} that $s \geq \frac{k-2}{7}$. Then, by considering the case $d=2$,
\beq
\begin{aligned}
\Sigma_s &\geq \frac{2s}{5} \sum_{d=1}^{\ceil*{\kappa(s)}-1} \frac{d\,\Omega_d}{k-d+1}\\
& \geq \frac{2s}{5(k-1)}\\
& \geq \frac{2(k-2)}{35(k-1)} \\
&\geq \frac{1}{35}
\end{aligned}
\eeq
for $k \geq 3$. Thus, if $n \geq 35k$,
\beq
\binom{k}{s} e^{-n\Sigma_s} \leq 2^k e^{-k}.
\eeq

\item If $\ceil*{\kappa(s)} = 1, 2$,
\beq
\Sigma_s \geq \frac{1}{5} \sum_{d=\ceil*{\kappa(s)}}^{k-\ceil*{\kappa(s)}} \Omega_d \geq \frac{\Omega_2}{5} = \frac{1}{10}.
\eeq
Thus, if $n \geq 10k$,
\beq
\binom{k}{s} e^{-n\Sigma_s} \leq 2^k e^{-k}.
\eeq
\end{enumerate}

\section{Proof of Lemma \ref{lem:rL}}\label{app:lem:rL}
In this lemma, we show that for $L=k, k-1, \dots, \delta k$, we have
\beq
	r(L) \geq \frac{C_e}{k}
\eeq
for a constant $\frac{1}{2} < C_e < 1-e^{-1}$.

By definition,
\beq
	r(k) = r(1, k) = \Omega_1 = \frac{1}{D}=\delta \geq \frac{1}{k}.
\eeq

If $d > D$, then $r(d, L) = \Omega_d \cdot q(d, L) = 0$, which makes the analysis different from Proposition 10 of \cite{luby2002lt}. However, if $d > k-L+1$, then $q(d, L) = 0$ anyway, and we can use Proposition 10 of \cite{luby2002lt} to find that
\beq
	r(L) = \frac{1}{k} \qquad \text{for } L \geq k-D+1.
\eeq
If $\delta k \geq k-D+1$, this proves the desired lemma. However, in our case $k-D \gg \delta k$ since $D = \frac{1}{\delta} < k^{\frac{1}{2} - \epsilon}$ and $\delta k \ll k$.
To find the lower bound on $r(L)$ for $L= k-D, \dots, \delta k$, we use the following lemma.

\begin{lemma} \label{lem:induction}
For any positive integers $a \geq b$,
\beq
	\frac{b}{a} + \frac{b(b-1)}{a(a-1)} + \dots + \frac{b(b-1) \cdots 1}{a(a-1) \cdots (a-b+1)} = \frac{b}{a-b+1}.
\eeq
\end{lemma}

\begin{IEEEproof}
For any fixed $c = a-b \geq 0$, we prove the lemma by induction on $b$. For $b=1$, it obviously holds since both sides are equal to $\frac{1}{a}$. Suppose that the formula holds for $b = b_0$. Then, for $b= b_0 +1$, $a=b+c = b_0+c+1$ and
\beq \begin{split}
	&\frac{b_0+1}{b_0+c+1} + \frac{(b_0+1)b_0}{(b_0+c+1)(b_0+c)} + \dots + \frac{(b_0+1)b_0(b_0-1) \cdots 1}{(b_0+c+1)(b_0+c)(b_0+c-1) \cdots (c+1)} \\
	&= \frac{b_0+1}{b_0+c+1} \left( 1 + \frac{b_0}{b_0+c} + \frac{b_0(b_0-1)}{(b_0+c)(b_0+c-1)} + \dots + \frac{b_0(b_0-1) \cdots 1}{(b_0+c)(b_0+c-1) \cdots (c+1)} \right) \\
	&= \frac{b_0+1}{b_0+c+1} \left( 1 + \frac{b_0}{c+1} \right) = \frac{b_0+1}{c+1}.
\end{split} \eeq
This proves the desired lemma.
\end{IEEEproof}

We now compute $r(L)$ for $L= k-D, \dots, \delta k$. Since
\beq
	\sum_{d=2}^{k-L+1} \frac{1}{d(d-1)} q(d, L) = \frac{1}{k},
\eeq
we find that
\beq \begin{split}
	&r(L) = \sum_{d=2}^{k-L+1} r(d, L) = \sum_{d=2}^{k-L+1} \frac{1}{d(d-1)} q(d, L) - \sum_{d=D+1}^{k-L+1} \frac{1}{d(d-1)} q(d, L) \\
	&= \frac{1}{k} - \frac{L(k-L-1)(k-L-2) \cdots (k-L-D+1)}{k(k-1)(k-2) \cdots (k-D)} - \cdots - \frac{L(k-L-1)(k-L-2) \cdots 1}{k(k-1)(k-2) \cdots L} \\
	&= \frac{1}{k} - \frac{L(k-L-1) \cdots (k-L-D+1)}{k(k-1) \cdots (k-D)} \left( 1 + \frac{k-L-D}{k-D-1} + \dots + \frac{(k-L-D) \cdots 1}{(k-D-1) \cdots L} \right) \\
	&= \frac{1}{k} - \frac{L(k-L-1)(k-L-2) \cdots (k-L-D+1)}{k(k-1)(k-2) \cdots (k-D)} \left( 1 + \frac{k-L-D}{L} \right),
\end{split} \eeq
where we used Lemma \ref{lem:induction} to get the last line. Thus,
\beq \begin{split}
	r(L) &= \frac{1}{k} \left( 1 - \frac{(k-L-1)(k-L-2) \cdots (k-L-D+1)}{(k-1)(k-2) \cdots (k-D+1)} \right) \\
	&= \frac{1}{k} \left( 1 - \left( 1- \frac{L}{k-1} \right) \left( 1 - \frac{L}{k-2} \right) \cdots \left( 1 - \frac{L}{k-D+1} \right) \right).
\end{split} \eeq
This in particular shows that $r(L)$ is an increasing function of $L$. Thus, we get
\beq
	r(L) \geq r(\delta k) = \frac{1}{k} \left( 1 - \left( 1- \frac{\delta k}{k-1} \right) \left( 1 - \frac{\delta k}{k-2} \right) \cdots \left( 1 - \frac{\delta k}{k-D+1} \right) \right).
\eeq
To estimate the right-hand side, we notice that
\beq
	\left( 1- \frac{\delta k}{k-1} \right) \left( 1 - \frac{\delta k}{k-2} \right) \cdots \left( 1 - \frac{\delta k}{k-D+1} \right) \leq \left( 1- \frac{\delta k}{k-1} \right)^D \to e^{-1}.
\eeq
In particular,
\beq
	r(L) \geq \frac{C_e}{k},
\eeq
for a constant $\frac{1}{2} < C_e < 1-e^{-1}$. 


\bibliographystyle{IEEEtran}

\end{document}